\documentclass[12pt]{article}
\pdfoutput=1
\usepackage[a4paper]{geometry}
\usepackage{jheppub, amsmath,amssymb,amsfonts,amsxtra, mathrsfs, makeidx,graphics,graphicx,amsthm,epsfig, ytableau,bm,longtable,float, color,tikz,mathtools,xfrac,footnote,rotating, lscape, makecell, environ,mathtools, empheq}
\usepackage{caption}

\usetikzlibrary{positioning}
\usetikzlibrary{chains}
\usetikzlibrary{arrows,fit,decorations.pathreplacing}
\tikzstyle{every picture}+=[remember picture]
\tikzstyle{na} = [baseline]

\usetikzlibrary{arrows, decorations.markings, calc, fadings, decorations.pathreplacing, patterns, decorations.pathmorphing, positioning}

\def\node#1#2{\overset{#1}{\underset{#2}{{\color{gray} \bullet}}}}

\def\node#1#2{\overset{#1}{\underset{#2}{\circ}}}

\tikzstyle{every picture}+=[remember picture]
\tikzstyle{na} = [baseline=-.5ex]

\usepackage{bbm}
\usepackage{subfigure}

\numberwithin{equation}{section}

\newcommand{\nn}{\nonumber}

\newcommand{\be}{\begin{equation}} \newcommand{\ee}{\end{equation}}
\newcommand{\bea}{\begin{equation} \begin{aligned}} \newcommand{\eea}{\end{aligned} \end{equation}}

\def\tilde{\widetilde}

\def\rt2{\sqrt{2}}

\def\det{\mathop{\rm det}}

\def\Tr{\mathop{\rm Tr}}

\def\CN{{\cal N}}

\def\CX{{\cal X}}
\def\CY{{\cal Y}}

\def\1{{\ds 1}}

\def\repa{\raise4pt\hbox{$\square$}\mkern-14mu\raise-4pt\hbox{$\square$}}
\def\repab{\overline{\raise4pt\hbox{$\square$}\mkern-14mu\raise-4pt\hbox{$\square$}\mkern-1mu}}

\def\smileface{\ensuremath{\hbox{\large$\bigcirc$}\mkern-15mu\raise-1pt\hbox{\scriptsize$\smallsmile$}%
\mkern-10mu\raise4pt\hbox{..}\mkern4mu}}
\def\frownface{\ensuremath{\hbox{\large$\bigcirc$}\mkern-15mu\raise-1pt\hbox{\scriptsize$\smallfrown$}%
\mkern-10mu\raise4pt\hbox{..}\mkern4mu}}

\newcommand{\ba}{\begin{array}}
\newcommand{\ea}{\end{array}}
\newcommand{\bi}{\begin{itemize}}
\newcommand{\ei}{\end{itemize}}

\def\bea#1\eea{\allowdisplaybreaks \begin{align}#1\end{align}}
 \newcommand{\ben}{\begin{enumerate}}
\newcommand{\een}{\end{enumerate}}
\newcommand{\bean}{\begin{eqnarray*}}
\newcommand{\eean}{\end{eqnarray*}}

\newcommand{\tr}{\mathrm{Tr}}

\newcommand{\W}{\mathcal{W}}

\newcommand{\comment}[1]{}

\definecolor{light-gray}{gray}{0.7}

\def\aup#1 {\overset{#1}{\uparrow} \, \overset{\tilde{#1}}{\downarrow}}

\title{Dualities for adjoint SQCD in three dimensions and emergent symmetries}
\author{Simone Giacomelli}

\affiliation{Mathematical Institute, University of Oxford, Andrew Wiles Building\\ Radcliffe Observatory Quarter (550), Woodstock Road,\\ Oxford, OX2 6GG, United Kingdom}
\emailAdd{simone.giacomelli@maths.ox.ac.uk}

\abstract{In this paper we study dualities for $\mathcal{N}=2$ gauge theories in three dimensions with matter in the fundamental and adjoint representation. The duality we propose, analogous to mirror symmetry, is obtained starting from $\mathcal{N}=4$ mirror theories and turning on a certain superpotential deformation involving monopole operators. We study the role of emergent symmetries in the dual theory, focusing on the case of models with gauge symmetry $U(2)$ or $SU(2)$. We find that $SU(2)$ adjoint SQCD with one flavor and zero superpotential is dual to SQED with two flavors and three singlets. As a byproduct, we recover several dualities for theories with $\mathcal{N}=2$ and $\mathcal{N}=4$ supersymmetry, including the duality appetizer of Jafferis and Yin.}

\begin{document}
\maketitle
\captionsetup[figure]{margin=10pt,font=small,labelfont=it,textfont={it}}

\section{Introduction}

Infrared dualities have been playing a central role over the last twenty years in the study of the dynamics of field theories in various dimensions, especially in the case of supersymmetric gauge theories. In this paper we will be concerned with infrared dualities for gauge theories in three dimensions with (at least) $\CN=2$ supersymmetry. Many examples of infrared dualities for this type of theories are known and in this paper we will be mainly concerned with mirror symmetry for $\CN=4$ theories and its generalization to models with $\CN=2$ supersymmetry \cite{Intriligator:1996ex, Hanany:1996ie, deBoer:1996mp, Porrati:1996xi, deBoer:1996ck, deBoer:1997ka, Aharony:1997bx, Assel:2014awa, Benvenuti:2016wet, Giacomelli:2017vgk, Fazzi:2018rkr}. 

These theories exhibit many remarkable properties which make them particularly interesting. First of all, they include a holomorphic sector protected against quantum corrections which provides a useful handle for the study of the theory. Another important property is that in 3d all gauge theories are asymptotically free (contrary to the 4d case) and therefore exhibit interesting dynamics, regardless of the choice of gauge group and matter content. Another interesting fact is the existence of monopole operators (for the definition see \cite{Aharony:1997bx} and also \cite{Chester:2017vdh, Assel:2018wtj} for an extensive discussion). These are local gauge invariant operators and therefore can be added to the lagrangian, although the resulting theory is not easy to interpret since these operators are not polynomial in the elementary fields. 

Models with monopole interactions frequently arise when one tries to compactify on $S^1$ a four dimensional theory and takes the 3d limit \cite{Aharony:2013dha}. On the other hand, sometimes it turns out that by turning on a monopole superpotential one flows to anther ``conventional" lagrangian theory \cite{Collinucci:2016hpz, Benini:2017dud, Collinucci:2017bwv} and in this case the monopole superpotential can be used as a tool to generate new infrared dualities (for other recent studies about monopole superpotentials see e.g. \cite{Amariti:2017gsm, Amariti:2018gdc, Amariti:2018dat, Aprile:2018oau, Amariti:2019rhc}). 

This is precisely the approach we follow in the present work to study $\CN=2$ theories with unitary gauge group and adjoint and fundamental matter fields (adjoint SQCD): starting from a carefully chosen $\CN=4$ gauge theory and turning on a suitable monopole superpotential one can flow, using the mechanism described in \cite{Benvenuti:2017kud}, to adjoint SQCD in the infrared. Knowing the mirror dual of the parent $\CN=4$ theory and implementing the same deformation in the dual theory we can derive a new duality, in the same spirit of \cite{Aharony:1997bx, Giacomelli:2017vgk}. This duality also admits a brane interpretation in Type IIB which is discussed in \cite{Benvenuti:2018bav}\footnote{See \cite{Benvenuti:2018bav, Amariti:2018wht} for a different duality of these models obtained via compactification of 4d dualities.}.

An interesting feature of our approach (based on the deformation of $\CN=4$ mirror theories) is that it can be used to systematically provide dual descriptions for models with an arbitrary number of adjoints and fundamentals, which is instead hard to achieve using the compactification method of  \cite{Benvenuti:2018bav, Amariti:2018wht} due to the constraints imposed on the matter content by asymptotic freedom in 4d. The price we have to pay, which constitutes the main focus of this note, is that the candidate dual theory is often plagued by emergent symmetries in the infrared (typically most of the symmetries which are not present in the parent $\CN=4$ theory). 

As we will see in Section \ref{intro}, in the dual description of adjoint SQCD (with 3 or more flavors) the infrared R-symmetry is invisible in the UV . This obstructs the computation of  the scaling dimension of chiral operators and makes it hard to understand in detail their mapping. As a result, this complicates the analysis of several relevant deformations in the dual theory.  

 In Section \ref{hiddenr} we will show that, at least in the case of $SU(2)$ (or $U(2)$) adjoint SQCD, the infrared R-symmetry can be recovered with a certain field redefinition (actually a duality for the underlying $\CN=4$ theories). After this modification, the mapping of chiral operators becomes easier: our dual of adjoint SQCD can be deformed to the known mirror duals of both $\CN=4$ and $\CN=2$ SQCD (hence can be considered the mirror dual of adjoint SQCD) and allows to understand the infrared properties of the so-called bad $\CN=4$ theories (in the language of \cite{Gaiotto:2008ak}) with simple field-theoretic manipulations (our results are in perfect agreement with the findings of \cite{Seiberg:1996nz, Assel:2017jgo, Dey:2017fqs, Assel:2018exy}). 

As a further test of our construction, in Section \ref{sect1} we use our proposal to recover the ``duality appetizer'' of \cite{Jafferis:2011ns}. In the process we will discover an abelian dual description for adjoint $SU(2)$ SQCD with one flavor and no superpotential. In the Appendix \ref{appdef} we analyze in detail the relevant deformations to $\CN=4$ and $\CN=2$ SQCD.

\section{Mirrors of N=2 theories}\label{intro} 

In this section we review the method developped in \cite{Giacomelli:2017vgk} to identify the mirror dual of $\CN=2$ SQCD and then generalize the construction to theories with adjoint matter.

\subsection{Mirror dual of SQCD}\label{sqcdrev} 

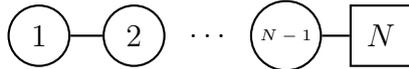
\begin{figure}
\begin{center}
\begin{tikzpicture}[->,thick, scale=0.5]
\node[circle, draw,  minimum height=8mm](L2) at (0.5,0){$1$};
\node[circle, draw,  minimum height=8mm](L3) at (3,0){$2$};
\node[](L4) at (5,0){$\dots$}; 
\node[circle, draw, inner sep=2.5](L5) at (7,0){{\tiny $N-1$}};
\node[rectangle, draw,  minimum width=6mm, minimum height=8mm,  minimum width=8mm](L6) at (9.5,0){$N$};

\draw[-] (L2) -- (L3);
\draw[-] (L5) -- (L6);

\end{tikzpicture} 
\end{center}
\caption{The $T(SU(N))$ theory. As is customary, a number $n$ inside a circle denotes a $U(n)$ gauge group and a line connecting two nodes a bifundamental hypermultiplet. The number inside a square denotes the number of hypermultiplets in the fundamental of the gauge group. We will use this notation throughout the paper. Unless otherwise specified, it should always be assumed (also when we discuss $\CN=2$ theories) that the spectrum includes a chiral multiplet in the adjoint representation for every gauge group. }\label{tsudis}
\end{figure}

The starting point in \cite{Giacomelli:2017vgk} is the following duality between $U(N)$ SQCD with $N+1$ flavors deformed by a monopole superpotential term and a WZ model found in \cite{Benini:2017dud}:  
\be \label{dualconfine}
\begin{array}{ccc}
\text{$U(N_c)$ with $N_f=N_c+1$} & \longleftrightarrow & \text{$N_f^2$ singlets $M$ and a singlet $\gamma$} \\
\text{ with $\W=\mathfrak{M}^+$} &  & \text{with $\W= \gamma\det(M)$}
\end{array}
\ee
where $\gamma$ is dual to the monopole $\mathfrak{M}^-$ in SQCD and $M$ is the counterpart of the meson $\widetilde{Q}_iQ^j$. For $N_c=1$ (\ref{dualconfine}) can also be extracted from mirror symmetry (see \cite{Collinucci:2016hpz}). This result is then used to prove that, by turning on a suitable monopole superpotential and repeatedly using (\ref{dualconfine}), the linear $\CN=4$ quiver usually called $T(SU(N))$ (see Figure \ref{tsudis}) can be converted into a single chiral multiplet in the adjoint of $SU(N)$.
 The precise statement is as follows: we start from $T(SU(N))$  then we deform the theory by adding singlets $\CX_1,\dots, \CX_{N-1}$ and turning on the following superpotential 
\be \label{WdefA1}
\begin{split}
\delta\W &=(\mathfrak{M}^{+00\cdots0}+ \mathfrak{M}^{0+0\cdots0}+ \mathfrak{M}^{00+\cdots0} + \ldots  + \mathfrak{M}^{000\cdots+})  \\
&\qquad + \CX_{1} [ \mathfrak{M}^{-00\cdots0}+ \mathfrak{M}^{0-0\cdots0}+ \mathfrak{M}^{00-\cdots0}+ \ldots (\text{terms with one minus})  ] \\
&\qquad +\CX_{2} [ \mathfrak{M}^{--0\cdots0}+ \mathfrak{M}^{0--\cdots0} +\ldots (\text{terms with two minuses}) ] + \ldots \\
&\qquad + \CX_{N-1} \mathfrak{M}^{---\cdots-}~,
\end{split}
\ee
where $\mathfrak{M}^{j_1 j_2 j_3 \cdots j_{N-1}}$ are the monopole operators carrying flux $(j_1, (j_2,0), \ldots, (j_{N-1},\dots,0))$ under $U(1)$, $U(2)$, $\cdots$, $U(N-1)$ gauge groups. In the infrared all the gauge nodes confine and the $SU(N)$ moment map turns into a free chiral multiplet in the adjoint of $SU(N)$. In the following we will refer to this procedure as "sequential confinement'' (see \cite{Benvenuti:2017kud} where this construction was introduced). 

\begin{figure}[h!]
\begin{center}
\begin{tikzpicture}[->,thick, scale=0.5]
\node[](L5) at (6,0){{\footnotesize $T(SU(N))$}};
\node[](L6) at (10,0) {{\footnotesize $SU(N)$}};
\node[rectangle, draw,  minimum width=6mm, minimum height=8mm,  minimum width=8mm](L7) at (13.5,0){$N+k$}; 

\node[rectangle, draw,  minimum width=6mm, minimum height=8mm,  minimum width=8mm](M1) at (19.5,0){$N$};
\node[circle, draw,  minimum height=8mm](M2) at (22,0){$N$};
\node[](M3) at (24,0){$\dots$}; 
\node[circle, draw,  minimum height=8mm](M4) at (26,0){$N$};
\node[](M5) at (29.5,0){{\footnotesize $T(SU(N))$}};
\node[circle, draw,  minimum height=8mm](M6) at (26,2.5){$1$};

\draw[-] (L5) -- (L6);
\draw[-] (L7) -- (L6);
\draw[=>] (15.5,0)--(18,0);
\draw[-] (M1) -- (M2);
\draw[-] (M4) -- (M5);
\draw[-] (M4) -- (M6);

\end{tikzpicture} 
\end{center}
\caption{The $\CN=4$ mirror pair used in \cite{Giacomelli:2017vgk} to extract the dual description of $\CN=2$ SQCD. The number of $U(N)$ gauge groups in the mirror quiver on the right is $k$.}\label{qcdmir}
\end{figure}
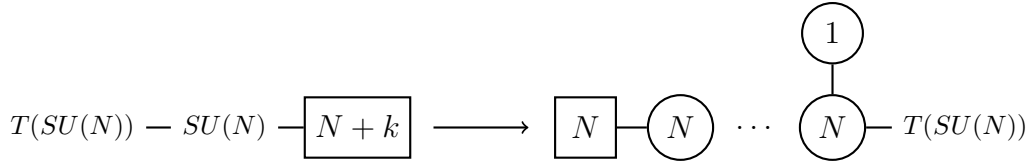

This observation is then used as follows: we start from $\CN=4$ $SU(N)$ SQCD coupled to $T(SU(N))$ and its mirror dual (see Figure \ref{qcdmir}).
The topological symmetry carried by the $T(SU(N))$ tail is mapped to the $SU(N)$ symmetry rotating the $N$ flavors in the mirror quiver. 
If we now turn on the monopole deformation (\ref{WdefA1}) for the $T(SU(N))$ theory on the left, it reduces at low energy to a chiral in the adjoint of $SU(N)$ (now gauged) which is coupled to the adjoint sitting in the $SU(N)$ vector multiplet. As a result, both adjoints can be integrated out leaving just $\CN=2$ SQCD without adjoint matter and no superpotential. Because of the $\CN=4$ mirror map, the  deformation (\ref{WdefA1}) is mapped in the mirror theory to a $\CX_i$-dependent mass matrix for the $N$ flavors and all of them except one become massive, leaving just one fundamental at low energy. More explicitly, the mass matrix reads 
\begin{equation}\label{massmat}
M=\left(\begin{array}{ccccc}
0 & 1 & 0 & \dots & 0 \\
\CX_{1} & 0 & 1 & \hphantom{X_{N}} & 0\\
\CX_{2} & \CX_{1} & \ddots & \ddots & \hphantom{X_{N}}\\
\vdots & \ddots & \ddots & 0 & 1\\
\CX_{N-1} & \dots & \CX_{2} & \CX_{1} & 0
\end{array}\right)\ .
 \end{equation} 
We therefore conclude that the mirror dual of $SU(N)$ SQCD with $N+k$ flavors is 
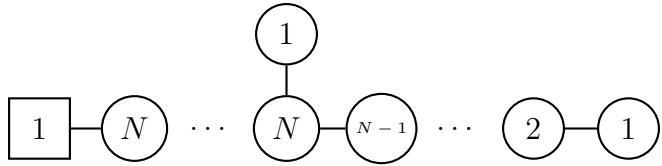
\begin{center}
\begin{tikzpicture}[->,thick, scale=0.5]
\node[rectangle, draw,  minimum width=6mm, minimum height=8mm,  minimum width=8mm](M1) at (19.5,0){$1$};
\node[circle, draw,  minimum height=8mm](M2) at (22,0){$N$};
\node[](M3) at (24,0){$\dots$}; 
\node[circle, draw,  minimum height=8mm](M4) at (26,0){$N$};
\node[circle, draw,  minimum height=8mm](M9) at (35,0){$1$};
\node[circle, draw,  minimum height=8mm](M8) at (32.5,0){$2$};
\node[](M7) at (30.5,0){$\dots$}; 
\node[circle, draw, inner sep=2.5](M5) at (28.5,0){{\tiny $N-1$}};
\node[circle, draw,  minimum height=8mm](M6) at (26,2.5){$1$};

\draw[-] (M1) -- (M2);
\draw[-] (M4) -- (M5);
\draw[-] (M4) -- (M6);
\draw[-] (M8) -- (M9);

\end{tikzpicture} 
\end{center}
The superpotential is the same one would write down for a $\mathcal{N}=4$ theory, except for terms involving the fundamental of $U(N)$ on the left (which we denote as $\tilde{q}$, $q$) which read (see \cite{Agarwal:2014rua} for the derivation)
\be\label{wsqcd}\W=\tilde{q}\phi^Nq+\sum_{i=1}^{N-1}\CX_{N-i}\tilde{q}\phi^{i-1}q+\dots\ee 
where $\phi$ indeed denotes the adjoint of the leftmost $U(N)$ gauge group. The global symmetry of $\CN=2$ $SU(N)$ SQCD with $N+k$ flavors has rank $2N+2k+1$ (including the $U(1)_R$ symmetry), whereas in the mirror quiver the manifest global symmetry has rank $N+k+2$. The emergence of a further $U(1)$ can be seen by applying the chiral ring stability criterion of \cite{Benvenuti:2017lle}, which implies that the first term in (\ref{wsqcd}) can be dropped. This allows for a $U(1)$ symmetry which acts on $\tilde{q}$, $q$ but not on $\phi$. This is identified with one of the Cartan generators of the axial $SU(N+k)$ symmetry. This fact is to be contrasted with the abelian case discussed in \cite{Aharony:1997bx}, in which the global symmetry groups of the dual theories manifestly have the same rank.  In any case, the most important point for the present work is that the infrared R-symmetry of the theory is manifestly visible in the mirror quiver. As we will see later, this is not the case for adjoint SQCD.

\subsection{Theory with adjoint matter}\label{adjsec} 

As we have seen, we can give mass and remove the adjoint chiral by coupling to the theory a $T(SU(N))$ tail and applying the sequential confinement procedure (i.e. turning on the monopole deformation (\ref{WdefA1})). Indeed, if we couple to the theory $n$ copies of $T(SU(N))$ and apply sequential confinement to all of them, we end up with $n+1$ adjoint chirals. Two of them become massive and at low energy we are left with $n-1$ adjoints and zero superpotential. From now on we will focus on the case $n=2$. 

\begin{figure}
\begin{center}
\begin{tikzpicture}[->,thick, scale=0.5]
\node[circle, draw, inner sep=2.5](L1) at (-10,0){ $1$};
\node[circle, draw, inner sep=2.5](L2) at (-7.5,0){ $2$};
\node[](L3) at (-5.5,0){$\dots$};
\node[circle, draw, inner sep=2.5](L4) at (-3.5,0){$N$};
\node[rectangle, draw, minimum width=6mm, minimum height=6mm](L5) at (-3.5,2.5){$k$};
\node[circle, draw, inner sep=2.5](L8) at (3,0){$1$};
\node[circle, draw, inner sep=2.5](L7) at (0.5,0){$2$};
\node[](L6) at (-1.5,0){$\dots$};
\node[](L9) at (-3.5,-2){\bf{\large{(A)}}};

\node[rectangle, draw,  minimum width=6mm, minimum height=6mm](B) at (18,0){$N$};
\node[circle, draw, inner sep=2.5](H3) at (15.5,0){$N$};
\node[](H2) at (13.5,0){$\dots$}; 
\node[circle, draw, inner sep=2.5](H1) at (11.5,0){$N$};
\node[rectangle, draw,  minimum width=6mm, minimum height=6mm](A) at (9,0){$N$};
\node[](H4) at (13.5,-2){\bf{\large{(B)}}};
 
\draw[-] (L1) -- (L2);
\draw[-] (L5) -- (L4);
\draw[-] (L7) -- (L8);
\draw[=>] (4.5,0)--(7.5,0); 
\draw[-] (H3) -- (B);
\draw[-] (A) -- (H1);

\end{tikzpicture} 
\end{center}
\caption{The mirror pair we will use to derive the dual of adjoint $U(N)$ SQCD. There are $k-1$ $U(N)$ gauge groups in the linear quiver on the right. The construction involves applying the sequential confinement procedure to the two $T(SU(N))$ tails on the left.} 
\label{mirradj}
\end{figure}
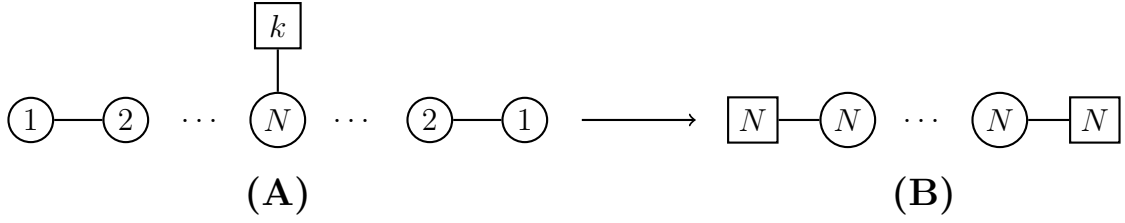

If we start from $\CN=4$ $U(N)$ SQCD with $k$ flavors and couple two $T(SU(N))$ tails, we find a theory whose mirror dual can be extracted using the brane construction of \cite{Hanany:1996ie}. The resulting mirror pair is as in Figure \ref{mirradj}. 

\noindent The quiver $(A)$ on the left has $SU(N)^2\times U(1)$ topological symmetry. The two $SU(N)$ factors are carried by the $T(SU(N))$ tail and correspond in the dual theory to the $SU(N)$ symmetries rotating the fundamentals at the ends of the quiver. As usual, the corresponding moment maps are related by the duality, in particular the monopole operators charged under the topological symmetry carried by each $T(SU(N))$ tail are mapped in the mirror theory to the "off-diagonal'' components of the mesons. 

\noindent When we turn on the monopole deformation (\ref{WdefA1}) for the two tails, both of them reduce to adjoint chirals and at low energy we are left with adjoint $U(N)$ SQCD with $k$ flavors (and a singlet we will discuss momentarily). As in the previous section, in the dual theory the monopole deformation is mapped to an off-diagonal mass term and, as a result, $N-1$ out of the $N$ flavors at each end become massive. We are therefore left with the candidate duality: 
\begin{figure}[h!]
\begin{center}
\begin{tikzpicture}[->,thick, scale=0.4]
\node[](L3) at (-5,0) {$\Phi$};
\node[] (L2) at (-0.3,0.7) {$\widetilde{Q}_i,Q_i$};
\node[circle, draw, inner sep=2.5](L4) at (-2.5,0){$N$};
\node[rectangle, draw, inner sep=1.7,minimum height=.6cm,minimum width=.6cm](L5) at (2,0){$k$};

\node[rectangle, draw, minimum height=.6cm,minimum width=.6cm](B) at (20.7,0){$1$};
\node[circle, draw, inner sep=2.5 ](H3) at (17,0){$N$};
\node[](H2) at (15,0){$\dots$}; 
\node[circle, draw, inner sep=2.5](H1) at (13,0){$N$};
\node[] (C) at (11,0.7) {$\tilde{q},q$};
\node[] (D) at (19,0.7) {$\tilde{p},p$};
\node[rectangle, draw, minimum height=.6cm,minimum width=.6cm](A) at (9.3,0){ $1$};
 
  \path[every node/.style={font=\sffamily\small,
  		fill=white,inner sep=1pt}]
(L4) edge [loop, out=145, in=215, looseness=4] (L4);
\draw[-] (L5) -- (L4);
\draw[=>] (4,0)--(7.5,0); 
\draw[-] (H3) -- (B);
\draw[-] (A) -- (H1);

\end{tikzpicture}
\end{center}
\caption{Adjoint $U(N)$ SQCD (here we indicate explicitly the adjoint with a loop) and its mirror dual.}\label{mirradj2}
\end{figure}

\noindent The superpotential is as in (\ref{wsqcd}): 
\be\label{wasqcd1}\W=\tilde{q}\phi_1^Nq+\sum_{i=1}^{N-1}\CX_{N-i}\tilde{q}\phi_1^{i-1}q+\dots+\tilde{p}\phi_{k-1}^Np+\sum_{i=1}^{N-1}\CY_{N-i}\tilde{p}\phi_{k-1}^{i-1}p,\ee 
where $\phi_1$ and $\phi_{k-1}$ denote the adjoint chirals of the leftmost and rightmost gauge groups in the figure respectively. We have suppressed all other superpotential terms, which are simply those of the parent $\CN=4$ theory. 

Since we are interested in $U(N)$ adjoint SQCD with zero superpotential, we need to manipulate the theory at hand a bit further: After the sequential confinement only the traceless part of the adjoint chiral in the $U(N)$ vectormultiplet acquires a mass. We therefore end up with adjoint SQCD plus one singlet (the trace part $\tr\Phi$ of the $U(N)$ adjoint) and superpotential $\W=\tr\Phi\widetilde{Q}_iQ_i$. Removing this term is easy: we introduce (as in \cite{Aharony:1997bx}) by hand a new chiral multiplet $S$ and we ``flip" $\tr\Phi$, meaning the new superpotential is 
$$\W=\tr\Phi\widetilde{Q}_iQ_i+S\tr\Phi.$$ 
Now both $S$ and $\tr\Phi$ become massive and can be integrated out. This procedure leaves at low energy $U(N)$ adjoint SQCD (with a traceless adjoint) with zero superpotential. This manipulation should of course be carried out in the dual quiver as well, therefore we couple $S$ to the dual counterpart of $\tr\Phi$. We can implement this by replacing (\ref{wasqcd1}) with 
\be\label{wasqcd}\W=\tilde{q}\phi_1^Nq+S\tilde{q}\phi_1^{N-1}q+\sum_{i=1}^{N-1}\CX_{N-i}\tilde{q}\phi_1^{i-1}q+\dots+\tilde{p}\phi_{k-1}^Np+\sum_{i=1}^{N-1}\CY_{N-i}\tilde{p}\phi_{k-1}^{i-1}p.\ee
Analogously to the model discussed in the previous section, the global symmetry of the theory (whose rank is $2k+2$) is not entirely visible in the mirror quiver: Including the two $U(1)$ factors we get (for $k>1$) by applying chiral ring stability \cite{Benvenuti:2017lle} (i.e. we drop from (\ref{wasqcd}) the terms $\tilde{q}\phi_1^Nq$ and $\tilde{p}\phi_{k-1}^Np$), the rank of the manifest global symmetry in the dual theory is $k+4$: The topological symmetry has rank $k-1$, there is an axial $U(1)$ symmetry acting on mesons and on adjoint chirals with opposite charge, there is also a $U(1)$ baryonic symmetry acting on fundamentals and antifundamentals with opposite charge and of course we have the UV $U(1)_R$ symmetry. However, contrary to the case of SQCD with fundamentals only, the infrared R-symmetry of adjoint SQCD is an hidden symmetry in the mirror theory, at least for $k\geq3$. This can be seen by considering monopole operators in the dual quiver, which are mapped to off-diagonal meson components in SQCD\footnote{This correspondence between monopole and mesons is simply inherited from the parent $\CN=4$ duality.}: by requiring all the monopole operators to have the same R-charge we find for $k\geq3$ the constraint 
\be\label{const}2R(q)+NR(\phi_1)=2R(p)+NR(\phi_{k-1})=2.\ee
In order to understand how this constraint arises (and its implications), it suffices to consider the case $k=3$. As we have explained, the dual theory is 
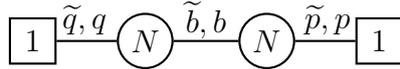
\begin{figure}[h!]
\begin{center}
\begin{tikzpicture}[->,thick, scale=0.4]
\node[rectangle, draw, minimum height=.6cm,minimum width=.6cm](B) at (20.7,0){$1$};
\node[circle, draw, inner sep=2.5 ](H3) at (17,0){$N$};
\node[circle, draw, inner sep=2.5](H1) at (13,0){$N$};
\node[] (C) at (11,0.7) {$\tilde{q},q$};
\node[] (D) at (19,0.7) {$\tilde{p},p$};
\node[] (E) at (15,0.7) {$\tilde{b},b$};
\node[rectangle, draw, minimum height=.6cm,minimum width=.6cm](A) at (9.3,0){ $1$};
 
\draw[-] (H3) -- (B);
\draw[-] (H3) -- (H1);
\draw[-] (A) -- (H1);

\end{tikzpicture}
\end{center} 
\end{figure}

\noindent If we denote by $r'$ the R-charge of the bifundamental $b$, then $R(\phi_1)=R(\phi_2)=2-2r'$. If we apply chiral ring stability to (\ref{wasqcd}), the R-charge of $q$ and $p$ is in principle unralated to that of $b$ and we denote it by $r$. We can now straightforwardly compute the R-charge of monopole operators $\mathfrak{M}^{+0}$, $\mathfrak{M^{0+}}$ and $\mathfrak{M^{++}}$, which are mapped to meson components and therefore should have the same R-charge. We find 
$$R(\mathfrak{M}^{+0})=R(\mathfrak{M}^{0+})=(r'-1)(N-2)+1-r;\quad R(\mathfrak{M}^{++})=(2N-2)(r'-1)+2-2r$$ 
and imposing their degeneracy we recover (\ref{const}), meaning that the extra $U(1)$ symmetries we gain from chiral ring stability do not mix with R-symmetry and are rather identified with two Cartan generators of the axial $SU(k)$ global symmetry. We therefore conclude that the R-symmetry is fixed up to a single unknown (say the R-charge of $q$) and is simply a combination of the UV R-symmetry and the axial $U(1)$ symmetry mentioned above. The trial R-symmetry in adjoint SQCD instead contains two unknowns: the charge of the fundamental flavors and the charge of the adjoint, which are not related by any symmetry argument. 

The conclusion is that in the mirror quiver we have automatically a constraint on R-charge assignments which is inherited from the parent $\CN=4$ theory. The interpretation is as follows: in the dual theory there are emergent symmetries in the infrared and the R-symmetry mixes with them. More precisely, the trial R-symmetry visible in the quiver corresponds, in adjoint SQCD, to a trial R-symmetry satisfying the constraint $R(\widetilde{Q}Q)=R(\Phi)$. This can be inferred by counting operators with the same trial R-charge in the quiver and comparing with the chiral ring of adjoint SQCD. For example, in the quiver we have the singlets $\CX_1$, $\CY_1$ and the quadratic Casimir of the adjoints $\Tr\phi_i^2$ which are all degenerate and uncharged under the topological symmetry. These can be matched with $\Tr\Phi^2$ and the diagonal components of the dressed meson $\widetilde{Q}_i\Phi Q^j$. As we will see shortly, it is also possible to show that $S^3_b$ partition functions of the two theories agree provided we set by hand $R(\widetilde{Q}Q)=R(\Phi)$ in adjoint SQCD. Since this constraint is not valid at the IR fixed point, we conclude that the $U(1)$ symmetry which assigns opposite charge to the mesons and to the adjoint chiral is not visible in the dual quiver. The purpose of the next Section is to show that this problem can be circumvented in the case $N=2$ with a field redefinition which makes the infrared R-symmetry manifestly visible. As we will see, our proposal passes several consistency checks. 

\subsection{Sphere partition functions}
 
The equivalence of $S^3_b$ partition functions (as defined in \cite{Hama:2010av, Hama:2011ea}) is proved using the same technique as in \cite{Giacomelli:2017vgk}, therefore we will be brief and refer the reader to that paper for notation and details. The derivation in  \cite{Giacomelli:2017vgk} builds on the equivalence of partition functions for the parent $\CN=4$ mirror theories. Actually, it is important to turn on the fugacity for the axial symmetry $H-C$ (the cartan generators of $SU(2)_C\times SU(2)_H$) which makes it impossible to explicitly compute the two partition functions and match them. The strategy is then to notice that the $\CN=4$ mirror theories of interest can be obtained by deforming $T(SU(n))$ theory with a suitable nilpotent vev for the HB (or CB for the mirror) moment map. The result then follows from the self-mirror property of $T(SU(n))$, which has been proven at the level of partition functions (with the fugacity for $H-C$ turned on) in \cite{Bullimore:2014awa}. 

In the case at hand we can use the same approach: we start from $T(SU(Nk))$, then we turn on for the CB moment map a nilpotent vev labelled by the partition $((k-1)^N,1^N)$ of $Nk$ (our convention is that the trivial vev is associated with the partition $(1^{Nk})$) and for the HB moment map a vev labelled by $(N^k)$. In this way the theory reduces to the quiver $(A)$ on the left of Figure \ref{mirradj}. By exchanging the roles of the two moment maps we get instead theory $(B)$ in Figure \ref{mirradj}. The rest of the argument is essentially as in \cite{Giacomelli:2017vgk}. In particular, when we monopole deform the two $T(SU(N))$ tails, their contribution in the partition function reduces to that of two $SU(N)$ adjoint chirals with the same R-charge as the meson. One of them cancels against the contribution from the adjoint in the $\CN=4$ vector multiplet and we are left with adjoint SQCD with $k$ flavors. The constraint $R(\widetilde{Q}Q)=R(\Phi)$ is then automatically satisfied. 

The $S_3^b$ partition function of theory $(A)$ in Figure \ref{mirradj} is 
\begin{eqnarray}\label{zundef}
\mathcal{Z}_{A}&=&\int\frac{\prod_{i=1}^{N}du_i}{N!}e^{2\pi i(\xi'+i\beta\frac{Q}{2})(\sum_iu_i)}\mathcal{Z}_{T(SU(N))}(u_i,\xi_i)\mathcal{Z}_{T(SU(N))}(u_i,z_i)\times \\ 
&& \frac{\prod_{i,j}s_b\left(u_i-u_j+m_A-i\frac{Q}{2}\alpha\right)\prod_{i=1}^{N}\prod_{j=1}^{k}s_b\left(i\frac{Q}{4}(1+\alpha)\pm u_i\mp m_j-\frac{m_A}{2}\right)}{\prod_{i<j}^{N}s_b\left(i\frac{Q}{2}\pm(u_i-u_j)\right)}\nonumber
\end{eqnarray}
where $\xi'$ is the FI parameter for the $U(N)$ gauge group and $\beta$ accounts for the mixing of the corresponding topological symmetry with the infrared R-symmetry. We have also introduced the real mass $m_A$ for $H-C$ and $\alpha$ denotes the corresponding mixing coefficient with the R-symmetry\footnote{$\alpha$ and $\beta$ can be determined via Z-extremization \cite{Jafferis:2010un}.}. $\mathcal{Z}_{T(SU(N))}$ denotes the contributions to the $S^3_b$ partition function from the $T(SU(N))$ tails and the parameters $\xi_i$, $z_i$ are the FI parameters of the corresponding topological symmetry.

\noindent It is now convenient to trade the FI parameters $\xi_i$, $\xi'$ and $z_i$ for $2N$ auxiliary parameters defined as follows: 
\be\xi_i=e_i-e_{i+1}\;\; (i=1\dots N);\quad \xi'=e_N-f_1;\quad z_i=f_i-f_{i+1}\;\; (i=1\dots N),\ee 
\be\label{constr} \sum_{i=1}^{N} e_i+(k-1)\sum_{i=1}^{N} f_i=0.\ee 
Equation (\ref{constr}) arises due to the nilpotent vev we are turning on for the CB moment map of $T(SU(Nk))$ (see the analogous discussion in \cite{Giacomelli:2017vgk} and especially \cite{Cremonesi:2014kwa} where this constraint is derived). 

In order to flow to adjoint SQCD we should now turn on the monopole superpotential described in Section \ref{sqcdrev}. The effect of this deformation is to break the topological symmetry of the $T(SU(N))$ tails and $H-C$ to the diagonal subgroup. In particular, we should identify all the parameters $\xi_i$, $z_i$ and $m_A$: 
\be\xi_i=z_i=m_A\equiv\xi+i\frac{Q}{2}\alpha.\ee 
The imaginary part accounts for the mixing with the IR R-symmetry. As a result, the parameters $e_i$ and $f_i$ defined before become 
\be\label{param1} e_i=\frac{k-1}{k}\left(\xi'+i\frac{Q}{2}\beta\right)+\left(\xi+i\frac{Q}{2}\alpha\right)\left(\frac{N+1-2i}{2}+\frac{(k-1)(N-1)}{k}\right);\ee 
\be\label{param2} f_i=-\frac{1}{k}\left(\xi'+i\frac{Q}{2}\beta\right)+\left(\xi+i\frac{Q}{2}\alpha\right)\left(\frac{N+1-2i}{2}-\frac{N-1}{k}\right).\ee 
Another result we need is the identity proven in \cite{Giacomelli:2017vgk}:  
\be\label{monid} \mathcal{Z}_{T^M(SU(N))}=e^{(N-1)\pi i(\xi+i\frac{Q}{2}(1-\alpha))(\sum_iu_i)}s_b^{N-1}\left(i\frac{Q}{2}\alpha-\xi\right)\prod_{i\neq j}s_b\left(u_i-u_j-\xi+i\frac{Q}{2}\alpha\right).\ee 
This states that after the monopole deformation $T(SU(N))$ reduces to a chiral multiplet in the adjoint of $SU(N)$. Plugging now (\ref{monid}) in (\ref{zundef}) we can easily see that, thanks to the identity $s_b(x)s_b(-x)=1$, the partition function of theory $(A)$ reduces to that of $U(N)$ SQCD with an adjoint (traceless) chiral whose R-charge is twice the R-charge of the fundamental matter and a singlet (as we have explained in Section \ref{adjsec}). We therefore see that the meson and the adjoint chiral have the same R-charge. 

The presence of the extra singlet we called $\tr\Phi$ in Section \ref{adjsec} can be seen as follows: the partition function of monopole deformed $T(SU(N))$ (\ref{monid}) does not cancel exactly against the contribution from the adjoint $U(N)$ chiral in (\ref{zundef}), leaving the term $s_b\left(\xi-i\frac{Q}{2}\alpha\right)$. We can remedy this by adding the flipping field $S$, which means multiplying (\ref{zundef}) by $s_b\left(i\frac{Q}{2}\alpha-\xi\right)$. Indeed, we should modify in the same way the partition function of theory $(B)$ as well. 

 The $S_3^b$ partition function of theory $(B)$ in Figure \ref{mirradj} is instead 
\begin{eqnarray}\label{zundef2}
 \mathcal{Z}_{B}=&\int\frac{\prod_{i=1}^{N}du_i}{N!}e^{2\pi i (m_1-m_2)(\sum_ju_j)} \frac{\prod_j\prod_{i=1}^{N}s_b\left(i \frac{Q}{4} \pm u_j \mp e_i +\frac{m_A}{2}\right) }{\prod_{i<j}^{N}s_b\left(i\frac{Q}{2}\pm(u_i-u_j)\right)}\times \nn \\
& \dots\int\frac{\prod_{i=1}^{N}dv_i}{N!}e^{2\pi i (m_{k-1}-m_k)(\sum_jv_j)} \frac{\prod_j\prod_{i=1}^{N}s_b\left(i \frac{Q}{4} \pm v_j \mp f_i +\frac{m_A}{2}\right) }{\prod_{i<j}^{N}s_b\left(i\frac{Q}{2}\pm(v_i-v_j)\right)}
\end{eqnarray} 
We have written explicitly only the contribution from the fundamentals at the left and right ends of the quiver $(B)$ and the Haar measure of the leftmost and rightmost $U(N)$ gauge groups. The dots denote all other terms. 

Since $T(SU(Nk))$ is self-mirror, we conclude that $$Z_A(m_A;e_i,f_i;m_j)=Z_B(-m_A;m_j;e_i,f_i),$$ where the parameters $e_i$ and $f_i$ are interpreted in theory $(B)$ as real masses for the fundamentals at the ends of the quiver and the parameters $m_j$ encode the FI parameters for the $k-1$ $U(N)$ gauge groups in the linear quiver. To conclude the argument, it suffices to notice that when we plug (\ref{param1}) and (\ref{param2}) in (\ref{zundef2}), the contribution of $N-1$ fundamentals at each end of the quiver cancel out (physically they become massive) thanks to the identity $s_b(x)s_b(-x)=1$. The duality therefore reduces to that of Figure \ref{mirradj2}.

\section{$U(2)$ and $SU(2)$ adjoint SQCD}\label{hiddenr} 

As we have argued before, the infrared R-symmetry of adjoint SQCD is a hidden symmetry in the dual theory (the quiver on the right in Figure \ref{mirradj2}). The goal of the present Section is to show that, in the case $N=2$, we can bypass this issue with a field redefinition. 

In order to explain how this works, let us start from the mirror pair discussed in \cite{Gaiotto:2008ak}: 
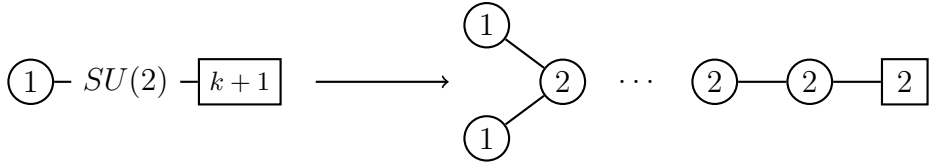
\begin{figure}[h!]
\begin{center}
\begin{tikzpicture}[->,thick, scale=0.5]
\node[circle, draw, inner sep=2.5](H1) at (-11,0){ $1$};
\node[](H2) at (-8.5,0){$SU(2)$};
\node[rectangle, draw, minimum width=6mm, minimum height=6mm](H3) at (-5.5,0){{\footnotesize $k+1$}};

\node[circle, draw, inner sep=2.5](L1) at (1,1.5){$1$};
\node[circle, draw, inner sep=2.5](L2) at (1,-1.5){$1$};
\node[circle, draw, inner sep=2.5](L3) at (3,0){$2$};
\node[](L4) at (5,0){$\dots$}; 
\node[circle, draw, inner sep=2.5](L5) at (7,0){$2$};
\node[circle, draw, inner sep=2.5](L6) at (9.5,0){$2$};
\node[rectangle, draw,  minimum width=6mm, minimum height=6mm](L7) at (12,0){ $2$};

\draw[-] (H1) -- (H2);
\draw[-] (H2) -- (H3);
\draw[=>] (-3.5,0)--(0,0); 
\draw[-] (L1) -- (L3);
\draw[-] (L2) -- (L3);
\draw[-] (L5) -- (L6);
\draw[-] (L6) -- (L7);

\end{tikzpicture} 
\end{center} 
\caption{On the left we have a special case of a $\mathcal{N}=4$ unitary quiver ending with a symplectic gauge group. The theory on the right is the corresponding mirror dual depicted in Figure 61 of \cite{Gaiotto:2008ak}.}\label{GWmir}
\end{figure}

\noindent The manifest $SO(2k+2)$ global symmetry acting on the $k+1$ $SU(2)$ doublets on the left arises in the dual theory due to the presence of monopole operators of dimension one. The $SU(2)$  topological symmetry associated with the $U(1)$ gauge node on the left corresponds instead to the symmetry rotating the two $U(2)$ doublets in the dual quiver. From this mirror pair we can obtain the following duality:
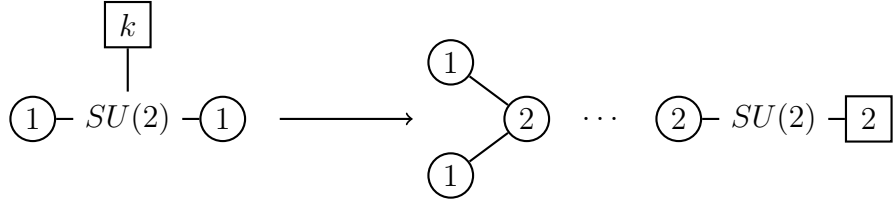
\begin{figure}[h!]
\begin{center}
\begin{tikzpicture}[->,thick, scale=0.5]
\node[circle, draw, inner sep=2.5](H1) at (-10,0){ $1$};
\node[](H2) at (-7.5,0){$SU(2)$};
\node[rectangle, draw, minimum width=6mm, minimum height=6mm](H3) at (-7.5,2.5){$k$};
\node[circle, draw, inner sep=2.5](H4) at (-5,0){$1$};

\node[circle, draw, inner sep=2.5](L1) at (1,1.5){$1$};
\node[circle, draw, inner sep=2.5](L2) at (1,-1.5){$1$};
\node[circle, draw, inner sep=2.5](L3) at (3,0){$2$};
\node[](L4) at (5,0){$\dots$}; 
\node[circle, draw, inner sep=2.5](L5) at (7,0){$2$};
\node[](L6) at (9.5,0){$SU(2)$};
\node[rectangle, draw,  minimum width=6mm, minimum height=6mm](L7) at (12,0){ $2$};

\draw[-] (H1) -- (H2);
\draw[-] (H2) -- (H3);
\draw[-] (H2) -- (H4);
\draw[=>] (-3.5,0)--(0,0); 
\draw[-] (L1) -- (L3);
\draw[-] (L2) -- (L3);
\draw[-] (L5) -- (L6);
\draw[-] (L6) -- (L7);

\end{tikzpicture} 
\end{center}
\caption{New $\mathcal{N}=4$ mirror pair obtained from the duality depicted in Figure \ref{GWmir} by gauging the $U(1)$ symmetry acting on one of the $k+1$ fundamentals.}\label{GWmir2}
\end{figure}

\noindent The only difference with respect to the previous case is that the $U(1)$ symmetry acting on one of the $k+1$ doublets has been gauged. As a result, we now have a $SU(2)^2\simeq SO(4)$ topological symmetry and a manifest $SO(2k)$ global symmetry. We therefore expect  the mirror theory to have a $SO(2k)$ topological symmetry and a $SO(4)$ global symmetry on the Higgs Branch. These are precisely the properties of the quiver on the right, which is obtained by  ``ungauging''\footnote{This is accomplished by gauging the $U(1)$ topological symmetry associated with the $U(2)$ gauge group, as explained in \cite{Witten:2003ya}.} the central $U(1)$ factor of the rightmost $U(2)$ gauge group in the quiver depicted in Figure \ref{GWmir}. The $SO(4)$ global symmetry acts on the two $SU(2)$ doublets. We therefore claim the quiver on the right of Figure \ref{GWmir2} is the corresponding mirror dual. 

Finally, with a further gauging, we can derive from the duality depicted in Figure \ref{GWmir2} the following mirror pair: 
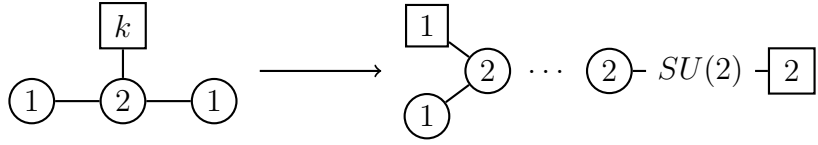
\begin{center}
\begin{tikzpicture}[->,thick, scale=0.4]
\node[rectangle, draw,  minimum width=5.5mm, minimum height=5.5mm](L1) at (5,1.5){$1$};
\node[circle, draw, inner sep=2.5](L2) at (5,-1.5){$1$};
\node[circle, draw, inner sep=2.5](L3) at (7,0){$2$};
\node[](L4) at (9,0){$\dots$}; 
\node[circle, draw, inner sep=2.5](L5) at (11,0){$2$};
\node[](L6) at (14,0){$SU(2)$};
\node[rectangle, draw,  minimum width=6mm, minimum height=6mm](L7) at (17,0){ $2$};

\node[circle, draw, inner sep=2.5](H1) at (-8,-1){ $1$};
\node[circle, draw, inner sep=2.5](H2) at (-5,-1){$2$};
\node[rectangle, draw, minimum width=6mm, minimum height=6mm](H3) at (-5,1.5){$k$};
\node[circle, draw, inner sep=2.5](H4) at (-2,-1){$1$};
 
\draw[-] (L1) -- (L3);
\draw[-] (L2) -- (L3);
\draw[-] (L5) -- (L6);
\draw[-] (L6) -- (L7);
\draw[=>] (-0.5,0)--(3.5,0); 
\draw[-] (H2) -- (H1);
\draw[-] (H2) -- (H3); 
\draw[-] (H2) -- (H4);

\end{tikzpicture}
\end{center}

\noindent Combining this with the mirror pair discussed in the previous section (see Figure \ref{mirradj}), we end up with the duality: 
\begin{figure}[h!]
\begin{center}
\begin{tikzpicture}[->,thick, scale=0.4]
\node[rectangle, draw,  minimum width=5.5mm, minimum height=5.5mm](L1) at (5,1.5){$1$};
\node[circle, draw, inner sep=2.5](L2) at (5,-1.5){$1$};
\node[circle, draw, inner sep=2.5](L3) at (7,0){$2$};
\node[](L4) at (9,0){$\dots$}; 
\node[circle, draw, inner sep=2.5](L5) at (11,0){$2$};
\node[](L6) at (14,0){$SU(2)$};
\node[rectangle, draw,  minimum width=6mm, minimum height=6mm](L7) at (17,0){ $2$};

\node[rectangle, draw,  minimum width=6mm, minimum height=6mm](B) at (-2,0){$2$};
\node[circle, draw, inner sep=2.5](H3) at (-4.5,0){$2$};
\node[](H2) at (-6.5,0){$\dots$}; 
\node[circle, draw, inner sep=2.5](H1) at (-8.5,0){$2$};
\node[rectangle, draw,  minimum width=6mm, minimum height=6mm](A) at (-11,0){$2$};
 
\draw[-] (L1) -- (L3);
\draw[-] (L2) -- (L3);
\draw[-] (L5) -- (L6);
\draw[-] (L6) -- (L7);
\draw[=>] (-0.5,0)--(3.5,0); 
\draw[-] (H3) -- (B);
\draw[-] (A) -- (H1);

\end{tikzpicture}
\end{center}
\caption{The $\mathcal{N}=4$ duality we use to construct the mirror dual of $U(2)$ adjoint SQCD.}\label{findual}
\end{figure}
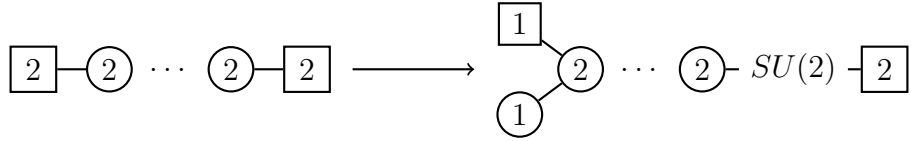

\noindent Notice that this is not a mirror pair: both theories have $(SU(2)\times SU(2)\simeq SO(4))\times U(1)$ global symmetry acting on the Higgs Branch and $SU(k)$ topological symmetry. 
As a simple consistency check of the duality, notice that for $k=2$ we recover a duality discussed in \cite{Collinucci:2017bwv}: The theory on the left becomes $U(2)$ SQCD with 4 flavors and the global symmetry enhances from $SU(2)^2\times U(1)$ to $SU(4)$. This enhancement is visible also on the right, since in this case the $SU(2)$ gauge group has three flavors and therefore $SO(6)\simeq SU(4)$ global symmetry. 

As is clear from the figure, we are basically redefining the $U(1)$ factors in the gauge group in such a way that the $SU(2)^2$ global symmetry acts on the two flavors on the right. As we will now see, this is convenient because turning on the nilpotent mass deformation in this duality frame allows to see explicitly an extra $U(1)$ factor, which is precisely the one we need to explicitly see the infrared R-symmetry. 

\subsection{Sequential confinement and its mirror}\label{supmass}

We find it more convenient to think of the two hypermultiplets in the fundamental of $SU(2)$ on the right of Figure \ref{findual} as a trifundamental half-hypermultiplet which we denote as $Q_{\alpha \beta \gamma}$. $SU(2)$ indices are contracted with the $\epsilon$ tensor and our convention will be that the first index denotes the gauged $SU(2)$. The other two are the $SU(2)^2$ flavor symmetry indices. 

The superpotential deformation we need to turn on in order to flow to the mirror dual of $U(2)$ adjoint SQCD (the mirror version of the monopole deformation describing sequential confinement) can be written as follows: 
\be\label{ccsup}\delta\W=\epsilon^{\alpha \beta}\epsilon^{\delta \gamma}[Q_{\alpha1\delta}Q_{\beta1\gamma}+Q_{\alpha\delta1}Q_{\beta\gamma1}+XQ_{\alpha2\delta}Q_{\beta2\gamma}+YQ_{\alpha\delta2}Q_{\beta\gamma2}],\ee 
where $X$ and $Y$ are chiral singlets. We clearly see from (\ref{ccsup}) that out of the four $SU(2)$ doublets two become massive and can be integrated out. The remaining massless fields are 
\be q_{\alpha}\equiv Q_{\alpha12}-Q_{\alpha21};\quad \tilde{q}^{\alpha}\equiv Q^{\alpha}_{22}.\ee 
In terms of $q$ and $\tilde{q}$ the superpotential can be written as follows: 
\be\label{supnew} \W=X'\tilde{q}q+\epsilon^{\alpha\beta}q_{\alpha}(\phi q)_{\beta}+Y'\epsilon_{\alpha\beta}(\tilde{q}\phi)^{\alpha}\tilde{q}^{\beta}+\dots\quad (X'\equiv Y-X;\;\; Y'\equiv X+Y)\ee 
where $\phi$ of course denotes the $SU(2)$ adjoint. We have suppressed superpotential terms not involving $q$ and $\tilde{q}$. We have also dropped terms involving higher powers of $\phi$ since they are not compatible with the chiral ring stability criterion of \cite{Benvenuti:2017lle}.

\begin{figure}[h!]
\begin{center}
\begin{tikzpicture}[->,thick, scale=0.4]
\node[](L3) at (-7,0) {$\Phi$};
\node[] (L2) at (-0.4,0.7) {$\widetilde{Q}_i,Q_i$};
\node[circle, draw, inner sep=2.5](L4) at (-3.3,0){$U(2)$};
\node[rectangle, draw, inner sep=1.7,minimum height=.8cm,minimum width=.8cm](L5) at (2,0){$k$};
\node[](W3) at (-2,-3.5) {{\bf\large{(A)}}};
\node[](W4) at (17,-3.5) {{\bf\large{(B)}}};

\node[rectangle, draw, minimum height=7mm,,minimum width=.7cm](A) at (9,2){$1$};
\node[] (H1) at (11.4,1.7) {$\tilde{p}_1,p_1$};
\node[circle, draw, minimum height=8mm](B) at (9,-2){$1$};
\node[] (H1) at (11.4,-1.7) {$\tilde{p}_2,p_2$};
\node[circle, draw, minimum height=8mm](C) at (12,0){$2$};
\node[] at (14,0) {$\dots$};
\node[circle, draw, minimum height=8mm](D) at (16,0){$2$};
\node[] (H4) at (18.2,0.7) {$\tilde{b}_1,b_1$};
\node[](C1) at (20.8,0){{\footnotesize $SU(2)$}};
\node[] (H3) at (22.9,0.7) {$\tilde{q},q$};
\node[rectangle, draw,  minimum width=8mm, minimum height=8mm](D1) at (25,0){ $1$};

  \path[every node/.style={font=\sffamily\small,
  		fill=white,inner sep=1pt}]
(L4) edge [loop, out=145, in=215, looseness=4] (L4);
\draw[-] (L5) -- (L4);
\draw[-] (C) -- (B);
\draw[-] (A) -- (C);
\draw[-] (C1) -- (D);
\draw[-] (C1) -- (D1);
\end{tikzpicture}
\end{center}
\caption{$U(2)$ adjoint SQCD (Theory A) with zero superpotential and its mirror dual (Theory B).}\label{mirru2}
\end{figure}
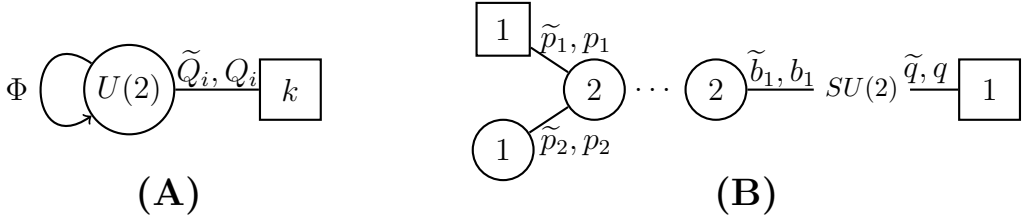

 In summary, the duality we are proposing is depicted in Figure \ref{mirru2}, 
where the dots stand for a linear quiver of $U(2)$ gauge groups with bifundamental hypermultiplets between them. In total there are $k-2$ $U(2)$ gauge groups and the same number of bifundamental hypers $\tilde{b}_i,b_i$. In the figure we included explicitly only the $SU(2)\times U(2)$ bifundamental $\tilde{b}_1,b_1$.
The superpotential indeed includes the terms in (\ref{supnew}). All other fields enter the superpotential via the standard $\CN=4$ couplings except $\tilde{p}_1$ and $p_1$, which are also coupled to an extra singlet which we call $S$. The addition of this singlet is required to remove the trace part of the $U(2)$ adjoint chiral inherited from the parent $\CN=4$ theory. The full superpotential is therefore 
\be\label{supcomp} \W=X'\tilde{q}q+\epsilon^{\alpha\beta}q_{\alpha}(\phi q)_{\beta}+Y'\epsilon_{\alpha\beta}(\tilde{q}\phi)^{\alpha}\tilde{q}^{\beta}+S\tilde{p}_1p_1+\W_{\CN=4}.\ee
Notice that the adjoint multiplet $\Phi$ in Figure \ref{mirru2} is traceless: we can neglect the trace part since it is a gauge singlet and decouples from the theory unless we turn on a superpotential interaction.

We can now understand why the field redefinition described above is useful: In the duality depicted in Figure \ref{mirradj2} the singlets $\CX_1$ and $\CY_1$ appearing in (\ref{wasqcd})  are on the same footing and therefore are expected to have the same R-charge. None of the symmetries (manifest in the lagrangian description) which assign different charge to $\CX_1$ and $\CY_1$ mix with the R-symmetry. This is to be contrasted with the corresponding fields $X'$ and $Y'$ appearing in (\ref{supcomp}), which do not need to have the same R-charge. We therefore see that the field redefinition described above allows to detect a $U(1)$ symmetry which is hidden in (\ref{wasqcd}) and this turns out to be precisely what we need to identify the correct infrared R-symmetry. We will give evidence for this claim below. 

\subsection{Duality for $SU(2)$ SQCD and study of the chiral ring} 

The dual description of $SU(2)$ adjoint SQCD with zero superpotential can be directly derived from the duality depicted in Figure \ref{GWmir2} upon turning on the relevant deformation described in Section \ref{supmass}:
\begin{figure}[h!]
\begin{center}
\begin{tikzpicture}[->,thick, scale=0.4]
\node[](L3) at (-7,0) {$\Phi$};
\node[] (L2) at (-0.4,0.7) {$\widetilde{Q}_i,Q_i$};
\node[circle, draw, inner sep=2.5](L4) at (-3.3,0){{\footnotesize $SU(2)$}};
\node[rectangle, draw, inner sep=1.7,minimum height=.8cm,minimum width=.8cm](L5) at (2,0){$k$};
\node[](W3) at (-2,-3.5) {{\bf\large{(A)}}};
\node[](W4) at (17,-3.5) {{\bf\large{(B)}}};

\node[circle, draw, minimum height=8mm](A) at (9,2){$1$};
\node[] (H1) at (11.4,1.7) {$\tilde{p}_1,p_1$};
\node[circle, draw, minimum height=8mm](B) at (9,-2){$1$};
\node[] (H1) at (11.4,-1.7) {$\tilde{p}_2,p_2$};
\node[circle, draw, minimum height=8mm](C) at (12,0){$2$};
\node[] at (14,0) {$\dots$};
\node[circle, draw, minimum height=8mm](D) at (16,0){$2$};
\node[] (H4) at (18.2,0.7) {$\tilde{b}_1,b_1$};
\node[](C1) at (20.8,0){{\footnotesize $SU(2)$}};
\node[] (H3) at (22.9,0.7) {$\tilde{q},q$};
\node[rectangle, draw,  minimum width=8mm, minimum height=8mm](D1) at (25,0){ $1$};

  \path[every node/.style={font=\sffamily\small,
  		fill=white,inner sep=1pt}]
(L4) edge [loop, out=145, in=215, looseness=4] (L4);
\draw[-] (L5) -- (L4);
\draw[-] (C) -- (B);
\draw[-] (A) -- (C);
\draw[-] (C1) -- (D);
\draw[-] (C1) -- (D1);
\end{tikzpicture}
\end{center}
\caption{$SU(2)$ adjoint SQCD  and its mirror dual.}\label{mirrsu2}
\end{figure}
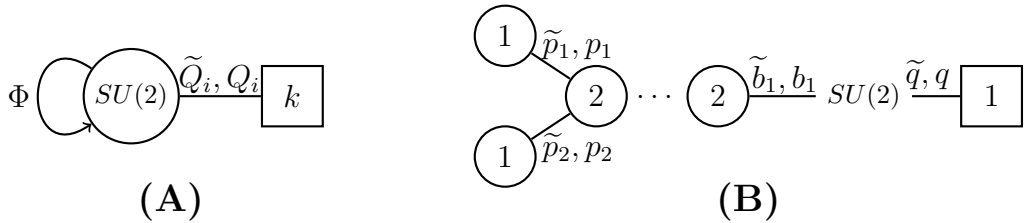
 The superpotential is exactly as in (\ref{supcomp}) (the term $S\tilde{p}_1p_1$ can be included in $\W_{\CN=4}$ since the $U(1)$ node is now gauged). 

\subsubsection{Mapping chiral operators}

As is customary in mirror symmetry, the meson components of $SU(2)$ SQCD are mapped to monopole operators (those with minimal magnetic charge under the $U(1)$ and/or $U(2)$ gauge groups only) in the dual theory and to  the (trace part of) chirals in the adjoint representation of the $U(1)$, $U(2)$ gauge groups in the dual theory. The monopoles of SQCD are instead mapped to chains of bifundamentals in the mirror quiver. More precisely, we propose the following map between chiral operators (see Figure \ref{mirrsu2}): 
\be\label{chimap}\begin{array}{c|c} 
\text{Theory A} & \text{Theory B} \\
\hline 
\tr\Phi^2 & Y' \\
\widetilde{Q}_i\Phi Q^i & X' \\ 
\mathfrak{M} & (p_1\dots b_1\tilde{q})(\tilde{q}\tilde{b}_1\dots\tilde{p}_1)\\
\{\mathfrak{M}\Phi\} & (p_1\dots b_1q)(\tilde{q}\tilde{b}_1\dots\tilde{p}_1)\\
\end{array}\ee
where $\{\mathfrak{M}\Phi\}$ denotes the dressed monopole operator \cite{Cremonesi:2013lqa}. The chains of bifundamentals inside each bracket are charged under the $U(1)$ gauge symmetry acting on $\tilde{p}_1,p_1$. Indeed, we could have considered the analogous operators with insertions of $\tilde{p}_2,p_2$, but exploiting F-terms one can check these are just equivalent in the chiral ring to the ones we have chosen. Analogously, the operator with two insertions of $q$ (the operator $(p_1\dots b_1q)(q\tilde{b}_1\dots\tilde{p}_1)$) is not an independent generator: The F-term for $\phi$ (see (\ref{supcomp})) implies this is equivalent in the chiral ring to $-Y' (p_1\dots b_1\tilde{q})(\tilde{q}\tilde{b}_1\dots\tilde{p}_1)$. All other chains of bifundamentals are trivial in the chiral ring or related to the ones discussed above. 

Our identification of $\widetilde{Q}_i\Phi Q^i$ with $X'$ can be tested as follows: if we add this operator to the superpotential the theory becomes $SU(2)$ $\CN=4$ SQCD and therefore we expect the mirror theory to reduce to the known mirror dual discussed in \cite{Hanany:1996ie}. This is precisely what happens if we deform the superpotential (\ref{supcomp}) by adding a term linear in $X'$, as we show in detail in Appendix \ref{appdef}. Analogously, if we add a mass term for the adjoint chiral we flow in the IR to $\CN=2$ SU(2) SQCD and therefore we can test our identification of $\tr\Phi^2$ with $Y'$ by checking that deforming (\ref{supcomp}) with a term linear in $Y'$ we recover the mirror dual proposed in \cite{Giacomelli:2017vgk} (see Appendix \ref{appdef} for the details).

Let us now discuss the R-charge of the theory. If we assign R-charge $r$ to the fundamentals $\widetilde{Q}_i, Q_i$ and $r'$ to the adjoint $\Phi$ of $SU(2)$ SQCD\footnote{The actual values of $r$ and $r'$ can be determined using Z-extremization.}, we should assign charge $1-r$ to the bifundamentals $\tilde{b}_i, b_i$ if we want the monopole operators to have R-charge $2r$ as the mesons of  $SU(2)$ SQCD. We should also assign charge $2r'$ to $Y'$ since it maps to $\tr\Phi^2$. From the superpotential (\ref{supcomp}) we then conclude that $q$ has charge $1-r$ and $\tilde{q}$ has charge $1-r-r'$. The charge of $X'$ is then $2r+r'$, which of course agrees with the R-charge of $\widetilde{Q}_i\Phi Q^i$. The R-charge of the monopole and dressed monopole operators of SQCD are then fixed to be $2k(1-r)-2r'$ and $2k(1-r)-r'$ respectively, in perfect agreement with those of the corresponding bifundamental chains appearing in (\ref{chimap}). 

The last aspect we need to discuss is the mapping of "dressed mesons'' $\widetilde{Q}_i\Phi Q^j$, whose R-charge is $2r+r'$. Let's consider the case of $U(2)$ SQCD which is simpler and will suffice to illustrate this point. 
As we have seen, the trace part is mapped to the singlet $X'$ in the mirror theory, so we need to discuss the remaining $k^2-1$ independent components. First of all, we can notice that the monopole operator with minimal magnetic charge under the $SU(2)$ gauge group in the mirror quiver has precisely R-charge $2r+r'$, suggesting that it might correspond to one of the components of the dressed meson. The problem is then reduced to the counting of monopole operators with charge $2r+r'$. 

To do this, we can notice that the mirror dual of $U(2)$ SQCD depicted in Figure \ref{mirru2} is essentially a $\CN=4$ unitary linear quiver ending with a symplectic gauge group\footnote{The presence of the singlet $S$ in (\ref{supcomp}), which spoils $\CN=4$ supersymmetry, is irrelevant for the purpose of computing the R-charge of monopole operators.} and therefore we can exploit the analysis performed in Section 5.3 of \cite{Gaiotto:2008ak}. The result is that all the monopole operators with R-charge $2r+r'$ have minimal magnetic charge under $SU(2)$. They can also have nontrivial magnetic charge of the form (up to permutation) $(1,0)$, $(0,-1)$ or $(1,-1)$ under the unitary groups. The rule is that the subquiver formed by the nodes at which the magnetic charge is $(1,m)$ (with $m$ either $0$ or $-1$) has to be connected. Analogously, nodes with magnetic charge $(m,-1)$ (with $m$ either 1 or 0) should form a connected subquiver. We stress that both subquivers contain the $SU(2)$ node. Notice that our mirror quiver terminates with a $U(1)$ node and the corresponding magnetic charge can only be $\pm1$ or $0$. We find a total of $k^2-1$ monopole operators with R-charge $2r+r'$ as desired and they transform in the adjoint representation of the $SU(k)$ topological symmetry supported by the unitary linear quiver as expected. 

\subsubsection{A chiral ring relation for $SU(2)$ adjoint SQCD}

 From our duality we can infer the following chiral ring relation for $SU(2)$ adjoint SQCD: 
\be\label{chiralrel}\{\mathfrak{M}\Phi\}^2=-\tr\Phi^2\mathfrak{M}^2.\ee 
In order to simplify the equations let's define (see (\ref{chimap}))
$$\mathcal{P}_{\alpha}\equiv (p_1\dots b_1)_{\alpha};\quad \widetilde{\mathcal{P}}_{\beta}\equiv(\tilde{b}_1\dots\tilde{p}_1)_{\beta},$$ 
where $\alpha$ and $\beta$ denote $SU(2)$ indices. Using now the identity $\epsilon^{\alpha\beta}\epsilon^{\gamma\delta}=\epsilon^{\alpha\gamma}\epsilon^{\beta\delta}-\epsilon^{\alpha\delta}\epsilon^{\beta\gamma}$, we find the relation 
$$\{\mathfrak{M}\Phi\}\simeq (\mathcal{P}q)(\tilde{q}\widetilde{\mathcal{P}})=\mathcal{P}_{\alpha}q_{\beta}\tilde{q}_{\gamma}\widetilde{\mathcal{P}}_{\delta}\epsilon^{\alpha\beta}\epsilon^{\gamma\delta}= \mathcal{P}_{\alpha}q_{\beta}\tilde{q}_{\gamma}\widetilde{\mathcal{P}}_{\delta}\epsilon^{\alpha\gamma}\epsilon^{\beta\delta}=(\mathcal{P}\tilde{q})(q\widetilde{\mathcal{P}}),$$ 
where we used the fact that $q_{\beta}\tilde{q}_{\gamma}\epsilon^{\beta\gamma}$ is set to zero by the F-term for $X'$ in (\ref{supcomp}). From this identity we therefore find 
$$\{\mathfrak{M}\Phi\}^2\simeq (\mathcal{P}q)(\tilde{q}\widetilde{\mathcal{P}})(\mathcal{P}q)(\tilde{q}\widetilde{\mathcal{P}})=(\mathcal{P}q)(q\widetilde{\mathcal{P}})(\mathcal{P}\tilde{q})(\tilde{q}\widetilde{\mathcal{P}})=-Y'(\mathcal{P}\tilde{q})(\tilde{q}\widetilde{\mathcal{P}})(\mathcal{P}\tilde{q})(\tilde{q}\widetilde{\mathcal{P}}),$$ 
and from (\ref{chimap}) we see that the r.h.s. is identified with $-\tr\Phi^2\mathfrak{M}^2$ in adjoint SQCD. Here we have exploited the chiral ring relation discussed before $(p_1\dots b_1q)(q\tilde{b}_1\dots\tilde{p}_1)=-Y' (p_1\dots b_1\tilde{q})(\tilde{q}\tilde{b}_1\dots\tilde{p}_1)$. 

\subsubsection{The chiral ring map for $U(2)$ adjoint SQCD} 

The above analysis can be repeated straightforwardly for $U(2)$ SQCD. The main difference is that now we have monopole operators with positive and negative topological charge. Chiral operators are mapped as follows: 
\be\label{chimap2}\begin{array}{c|c} 
\text{Theory A} & \text{Theory B} \\
\hline 
\tr\Phi^2 & Y' \\
\widetilde{Q}_i\Phi Q^i & X' \\ 
\widetilde{Q}_i Q^i & S \\
\mathfrak{M}^+ & p_1\dots b_1\tilde{q}\\
\mathfrak{M}^- & \tilde{q}\tilde{b}_1\dots\tilde{p}_1\\
\{\mathfrak{M}\Phi\}^{+} &  p_1\dots b_1q\\
\{\mathfrak{M}\Phi\}^{-} & q\tilde{b}_1\dots\tilde{p}_1\\
\end{array}\ee
Notice that now the product of bifundamentals $p_1\dots b_1\tilde{q}$ (and all the other analogous products appearing in (\ref{chimap2})) is gauge invariant since the $U(1)$ symmetry acting on $p_1$ is no longer gauged. Exploiting again the equation $(p_1\dots b_1q)(q\tilde{b}_1\dots\tilde{p}_1)=-Y' (p_1\dots b_1\tilde{q})(\tilde{q}\tilde{b}_1\dots\tilde{p}_1)$ we find the $U(2)$ counterpart of (\ref{chiralrel}): 
 \be\label{chiralrel2}\{\mathfrak{M}\Phi\}^+\{\mathfrak{M}\Phi\}^-=-\tr\Phi^2\mathfrak{M}^+\mathfrak{M}^-.\ee

\section{Adjoint SQCD with one flavor and the duality appetizer} \label{sect1}

In this section we will use our duality to study adjoint SQCD with one and zero fundamentals. We will see that the theory with $N_f=1$ and zero CS level is equivalent to an abelian theory. This will allow us to recover in a simple way the ``duality appetizer'' of \cite{Jafferis:2011ns}: the $SU(2)$ theory with an adjoint chiral and CS level one is equivalent to a free chiral plus a topological sector. We will also recover a duality recently proposed in \cite{Dimofte:2017tpi} for the same theory with zero CS level.   

\subsection{The abelian dual of adjoint SQCD with one flavor} 

In order to derive the dual description of $SU(2)$ adjoint SQCD with one flavor, we start from the duality for the theory with $N_f=2$: 
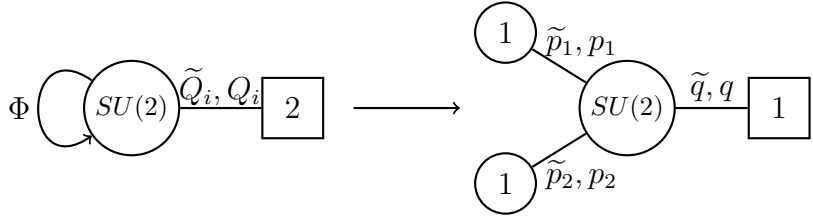
\begin{figure}[h!]
\begin{center}
\begin{tikzpicture}[->,thick, scale=0.4]
\node[](L3) at (-7,0) {$\Phi$};
\node[] (L2) at (-0.4,0.7) {$\widetilde{Q}_i,Q_i$};
\node[circle, draw, inner sep=2.5](L4) at (-3.3,0){{\footnotesize $SU(2)$}};
\node[rectangle, draw, inner sep=1.7,minimum height=.8cm,minimum width=.8cm](L5) at (2,0){$2$};

\node[circle, draw, minimum height=8mm](A) at (9,2.5){$1$};
\node[] (H1) at (11.5,2.2) {$\tilde{p}_1,p_1$};
\node[circle, draw, minimum height=8mm](B) at (9,-2.5){$1$};
\node[] (H1) at (11.5,-2.2) {$\tilde{p}_2,p_2$};
\node[circle, draw, inner sep=2.5](C) at (13,0){{\footnotesize $SU(2)$}};
\node[] (H3) at (15.8,0.7) {$\tilde{q},q$};
\node[rectangle, draw,  minimum width=8mm, minimum height=8mm](D) at (18,0){ $1$};

  \path[every node/.style={font=\sffamily\small,
  		fill=white,inner sep=1pt}]
(L4) edge [loop, out=145, in=215, looseness=4] (L4);
\draw[-] (L5) -- (L4);
\draw[=>] (4,0)--(7.5,0); 
\draw[-] (C) -- (B);
\draw[-] (A) -- (C);
\draw[-] (C) -- (D);
\end{tikzpicture}
\end{center}
\caption{$SU(2)$ adjoint SQCD with two flavors and its mirror dual.}\label{mirrsu22}
\end{figure}
and turn on a complex mass term for one of the flavors ($\delta\W=m\widetilde{Q}_1Q_1$). In the mirror theory this is mapped to a "complex FI term'' at the two abelian tails ($\delta\W=m\phi_1-m\phi_2$). This  induces an expectation value for the bifundamentals $p_i,\tilde{p}_i$ which in turn higgses the gauge group down to $U(1)$. Only the fields $q$ and $\tilde{q}$ remain massless and become two flavors (which we denote as $p$ and $q$) of the unbroken $U(1)$ gauge group. There are also three massless singlets and the superpotential of the resulting abelian theory is 
\be\label{abdual}\W=X'(\tilde{p}q+\tilde{q}p)+\varphi(Y'\tilde{p}p-\tilde{q}q),\ee 
where $\varphi$ is the linear combination of $\phi_1$, $\phi_2$ and the Cartan component of the $SU(2)$ adjoint chiral which remains massless.  

\noindent The map between chiral operators is as follows: 
\be\label{chicorr}\begin{array}{c|c|c} 
\text{$SU(2)$ SQCD} & \text{SQED with 2 flavors} & \text{Trial R-charge} \\
\hline 
\widetilde{Q}Q & \varphi & 2r\\
\tr\Phi^2 & Y' & 2r' \\
\widetilde{Q}\Phi Q & X' & 2r+r' \\ 
\epsilon^{ab}Q_a(\Phi Q)_b & \mathfrak{M}^+ & 2r+r' \\
\epsilon_{ab}(\widetilde{Q}\Phi)^a\widetilde{Q}^b & \mathfrak{M}^- & 2r+r' \\ 
\mathfrak{M} & \tilde{p}p & 2-2r-2r'\\
\{\mathfrak{M}\Phi\} & \tilde{p}q & 2-2r-r'\\
\end{array}\ee 
The chiral ring relation (\ref{chiralrel})
$$\{\mathfrak{M}\Phi\}^2=-\tr\Phi^2\mathfrak{M}^2$$ 
follows simply from F-terms in the dual abelian theory: 
$$ \tilde{p}q \tilde{p}q= -\tilde{p}q \tilde{q}p= -\tilde{p}p \tilde{q}q=-Y' (\tilde{p}p)^2.$$ 
In the first equality we have exploited the F-term for $X'$ and in the third the F-term for $\varphi$.

As a simple consistency check, we can make contact with the duality discussed in \cite{Benvenuti:2017kud}. To this end we have to turn on in the $SU(2)$ theory the superpotential 
\be\label{iobenve}\W=\alpha\widetilde{Q}Q+\beta\tr\Phi^2,\ee 
where $\alpha$ and $\beta$ are chiral singlets. In \cite{Benvenuti:2017kud} this model was argued to be dual to $\CN=4$ SQED with two flavors. Our duality is indeed perfectly consistent with this claim: mapping the two superpotential terms in the abelian theory we find 
$$\W=X'(\tilde{p}q+\tilde{q}p)+\varphi(Y'\tilde{p}p-\tilde{q}q)+\alpha\varphi+\beta Y',$$ 
and integrating out massive fields we are left at low energy with the superpotential 
$$\W=X'(\tilde{p}q+\tilde{q}p).$$ 
Modulo a field redefinition this is precisely the lagrangian of the $\CN=4$ theory. 

Actually, we can use our duality to clarify one aspect of the duality discussed in \cite{Benvenuti:2017kud} (see also \cite{Benvenuti:2017bpg}): in that paper it was argued that the chiral operator $\beta$ appearing in (\ref{iobenve}) is zero in the chiral ring due to a quantum chiral ring relation. We are now in the position to identify precisely this relation: if instead of (\ref{iobenve}) we turn on the second superpotential term ($\W=\beta\tr\Phi^2$) only, in the abelian theory we find 
$$\W=X'(\tilde{p}q+\tilde{q}p)+\varphi(Y'\tilde{p}p-\tilde{q}q)+\beta Y'.$$ 
We see that $\beta$ becomes massive and the F-term for $Y'$ imposes the chiral ring relation $\beta=\varphi'\tilde{p}p$, which in the original $SU(2)$ theory reads 
\be\beta=\mathfrak{M}\widetilde{Q}Q.\ee 
Now it is clear that turning on the other term $\alpha\widetilde{Q}Q$ sets to zero in the chiral ring the meson $\widetilde{Q}Q$ and therefore the above relation reduces to $\beta=0$.

It is not harder to analyze the theory with gauge group $U(2)$: we should gauge the baryon number, which in the abelian dual is mapped to the topological symmetry as is clearly displayed in (\ref{chicorr}). The net effect is therefore to ungauge the $U(1)$ and therefore the dual theory is simply a WZ model with 7 chirals and superpotential (\ref{abdual}). If we further turn on the superpotential term $\beta\Tr\Phi^2$ in the $U(2)$ theory (and therefore make $Y'$ massive) we get a dual theory involving 6 singlets and superpotential 
\be\label{abdual2}\W=X'(\tilde{p}q+\tilde{q}p)+\varphi\tilde{q}q.\ee 
The map between chiral operators can be easily derived from (\ref{chicorr}). This is in perfect agreement with the result recently found in \cite{Benvenuti:2018bav}. 

Another consistency check is obtained by giving mass to the adjoint, thus flowing in the IR to $U(2)$ SQCD with one flavor. On the dual side we should turn on a linear superpotential term for $Y'$, which induces a vev for $\varphi\tilde{p}p$. From (\ref{abdual}) it is easy to see that $X'$, $Y'$, $q$ and $\tilde{q}$ become massive and we recover the known result (see \cite{Aharony:1997bx}) that at low energy the $U(2)$ theory with $N_f=1$ is described by three chirals (the meson $M$ and the two monopoles $\mathfrak{M}^{\pm}$), satisfying the constraint 
$$\mathfrak{M}^+\mathfrak{M}^-M=1.$$ 

\subsection{The theory without fundamentals and the duality appetizer}

In order to flow to the $SU(2)$ theory without fundamental fields we can simply give mass to the flavor, either real or complex. 

\subsubsection{Real mass deformation and the duality appetizer}

Let's start by considering the first option: In the model with one flavor there is indeed an axial $U(1)$ symmetry under which both $Q$ and $\widetilde{Q}$ have charge $+1$. If we turn on a real mass for this symmetry, at low energy we are left with $SU(2)$ SYM with CS level 1 (see \cite{Aharony:1997bx}) and an adjoint chiral multiplet. In the dual abelian theory both flavors have charge $-1$ under this symmetry and the singlets $X'$ and $\varphi$ have charge $+2$, whereas the singlet $Y'$ is uncharged. Therefore, in the dual theory the effect of this deformation is to make all the fields massive except $Y'$, which decouples in the IR being uncharged under the gauge group. Moreover, when we integrate out the two flavors we generate a CS level $-2$. The low energy theory thus reduces to a free chiral multiplet (which corresponds to $\tr\Phi^2$ in the non abelian theory) plus a topological sector: a $U(1)$ theory without matter fields and CS level $-2$. This is precisely the duality appetizer of \cite{Jafferis:2011ns}. Notice that\footnote{This follows e.g. from the level-rank duality for $\CN=2$ CS theories \cite{Giveon:2008zn, Benini:2011mf}.} the topological $U(1)$ theories with CS level $\pm2$ are equivalent and therefore, despite the nontrivial CS level, the model we are discussing is actually parity invariant.

\subsubsection{Complex mass deformation and $SU(2)$ $\mathcal{N}=4$ SYM}

Let's now come to the analysis of the complex mass: in this case we do not generate a CS term and the low-energy theory is simply $SU(2)$ SYM with an adjoint chiral and zero superpotential. This model actually has enhanced supersymmetry (it is $\CN=4$ $SU(2)$ SYM) and was recently discussed in \cite{Dimofte:2017tpi}, with the conclusion that the theory is equivalent at low-energy to a free hypermultiplet. We can recover this conclusion by analysing the effect of the complex mass term in the dual abelian theory, whose superpotential becomes 
\be\label{abdual1}\W=X'(\tilde{p}q+\tilde{q}p)+\varphi(Y'\tilde{p}p-\tilde{q}q)+m\varphi.\ee 
Now the F-term for $\varphi$ forces the flavors to acquire a nontrivial expectation value. This spontaneously breaks the gauge symmetry and one combination of the matter fields recombines with the vectormultiplet into a long multiplet. By expanding the superpotential around the vev, we find that four out of the six surviving chirals become massive and we are left with two free chirals at low energy. These correspond to the monopole operator and $\tr\Phi^2$ in the $SU(2)$ theory, in agreement with the analysis in Section 10 of \cite{Dimofte:2017tpi}. 

It is not harder to study $SU(2)$ or $U(2)$ $\CN=4$ SQCD with one flavor, starting from (\ref{abdual}). We find that the low energy theory is free and is described by a single hypermultiplet for $SU(2)$ and two hypermultiplets for $U(2)$. This agrees with the analysis of \cite{Assel:2018exy}.

\section{Concluding remarks}

In this note we have studied adjoint SQCD and derived a dual description using sequential confinement. The method is systematic and in principle can be used to propose dualities for models with an arbitrary number of adjoint multiplets. As we have seen, a potential technical problem of this approach is the presence of accidental symmetries. 

However, with suitable field redefinitions we can identify a duality frame in which this issue is not severe (at least for gauge group $U(2)$ or $SU(2)$): The infrared R-symmetry is explicitly visible in the UV and the duality is suited as a starting point for studying the infrared dynamics of bad $\CN=4$ theories, for which standard techniques are harder to apply. We can also rederive with simple field-theoretic manipulations various known dualities, for theories with or without adjoint matter (see especially the discussion in Appendix \ref{appdef}).

One natural direction for future investigations is the study of theories with higher rank gauge groups. In this case we did not find a simple field redefinition analogous to that of Figure \ref{findual} and maybe one has to look at duality frames involving non lagrangian building blocks: In Figure \ref{findual} the $SU(2)$ vectormultiplet is coupled to two fundamentals and we can think of those as (the dimensional reduction of) $T_2$ (see \cite{Tachikawa:2015bga} for a review of $T_N$ theory), suggesting that in the general case a dual description involving $T_N$ might be the most convenient starting point. It would be interesting to understand emergent symmetries for this class of theories.

\acknowledgments{I would like to thank Sergio Benvenuti, Noppadol Mekareeya, Sara Pasquetti and Wolfger Peelaers for discussions. I would also like to thank the Galileo Galilei Institute for Theoretical Physics for hospitality during the completion of this
project. This work is supported by the ERC Consolidator Grant 682608 “Higgs bundles: Supersymmetric Gauge Theories and Geometry (HIGGSBNDL)”.}

\appendix 

\section{Superpotential deformation and comparison with known dualities}\label{appdef} 

In this Appendix we show in detail that the proposed mirror of $\CN=2$ adjoint SQCD can be deformed to the known mirrors of SQCD with eight and four supercharges respectively.

\subsection{Deformation to the $\CN=4$ theory} 

We start from the mirror dual of $SU(2)$ adjoint SQCD with $k>3$ flavors:
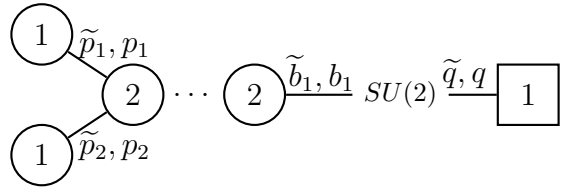
\begin{center}
\begin{tikzpicture}[->,thick, scale=0.4]
\node[circle, draw, minimum height=8mm](A) at (9,2){$1$};
\node[] (H1) at (11.4,1.7) {$\tilde{p}_1,p_1$};
\node[circle, draw, minimum height=8mm](B) at (9,-2){$1$};
\node[] (H1) at (11.4,-1.7) {$\tilde{p}_2,p_2$};
\node[circle, draw, minimum height=8mm](C) at (12,0){$2$};
\node[] at (14,0) {$\dots$};
\node[circle, draw, minimum height=8mm](D) at (16,0){$2$};
\node[] (H4) at (18.2,0.7) {$\tilde{b}_1,b_1$};
\node[](C1) at (20.8,0){{\footnotesize $SU(2)$}};
\node[] (H3) at (22.9,0.7) {$\tilde{q},q$};
\node[rectangle, draw,  minimum width=8mm, minimum height=8mm](D1) at (25,0){ $1$};

\draw[-] (C) -- (B);
\draw[-] (A) -- (C);
\draw[-] (C1) -- (D);
\draw[-] (C1) -- (D1);
\end{tikzpicture}
\end{center} 
According to our chiral ring map, in order to flow to the mirror of $\CN=4$ SQCD we should add a superpotential term linear in the singlet $X'$, therefore the superpotential becomes (see (\ref{supnew}))
\be\label{pots1}\W=X'\tilde{q}q-X'+\epsilon^{ab}q_a(\phi q)_b+Y'\epsilon_{ab}(\tilde{q}\phi)^a\tilde{q}^b+\Tr(\Psi b_1\tilde{b}_1)-\Tr(\phi\tilde{b}_1b_1)+\dots\ee
where $\Psi$ denotes the adjoint of the rightmost $U(2)$ gauge group in the figure. The other superpotential terms will not be relevant for our analysis. 

The new superpotential term induces a nonzero vev for $\tilde{q}q$ and, modulo a gauge transformation, we can solve D-terms by setting 
\be\tilde{q}=(\begin{array}{cc} 1 & 0 \end{array});\;\; q=\left(\begin{array}{c}1 \\ 0\end{array}\right)\ee 
this breaks the $SU(2)$ gauge group completely and when we expand (\ref{pots1}) around this vev we get 
\be\W=\Tr(\Psi b_1\tilde{b}_1)-\Tr(\phi\tilde{b}_1b_1)+X'(\tilde{q}\langle q\rangle+\langle\tilde{q}\rangle q)+\phi_{21}+Y'\phi_{12}+\dots\ee 
where $\phi_{ij}$ indeed denotes the component $(i,j)$ of the $SU(2)$ adjoint. We conclude that $X'$, $Y'$ and $\phi_{12}$ (together with one component of $q$, $\tilde{q}$) become massive. As we clearly see the expansion around the vev produces another linear superpotential term which in turn induces a vev for $(\tilde{b}_1b_1)_{12}$. The D-terms associated with the $U(2)$ gauge group are solved by setting\footnote{Our convention is that $b_1$ is a fundamental of $U(2)$ and $\tilde{b}_1$ an antifundamental.} 
\be\label{vevinter}\tilde{b}_1=\left(\begin{array}{cc} 1 & 0 \\ 0 & 0\end{array}\right);\quad b_1=\left(\begin{array}{cc} 0 & 1 \\ 0 & 0\end{array}\right)\ee 
and this vev breaks the $U(2)$ gauge group to the following $U(1)$ subgroup: 
\be\label{subgroup}\left(\begin{array}{cc} 1 & 0 \\ 0 & e^{i\alpha}\end{array}\right)\ee
Expanding the superpotential around this vev we find that all the components of $\phi$, $b_1$ and $\tilde{b}_1$ disappear from the low energy spectrum: either they recombine with the vectormultiplets into long multiplets or become massive. Also all the components of $\Psi$ except $\Psi_{22}$ become massive and are set to zero in the chiral ring. As a result the bifundamental $b_2$, $\tilde{b}_2$ (charged under the two rightmost $U(2)$ gauge groups) reduces to two doublets of the unbroken $U(2)$ and only one of them is charged under the $ U(1)$ gauge group (\ref{subgroup}). This doublet is also coupled to $\Psi_{22}$ which now plays the role of the chiral singlet sitting in the $U(1)$ $\CN=4$ vector multiplet. 

All in all, the low energy effective theory we are left with is described by the quiver
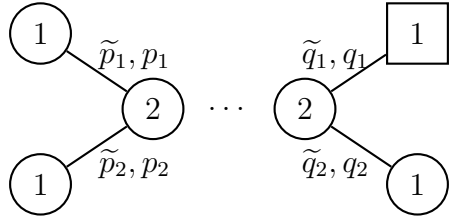
\begin{center}
\begin{tikzpicture}[->,thick, scale=0.4]
\node[circle, draw, minimum height=8mm](A) at (8.3,2.5){$1$};
\node[] (H1) at (11.4,1.8) {$\tilde{p}_1,p_1$};
\node[circle, draw, minimum height=8mm](B) at (8.3,-2.5){$1$};
\node[] (H1) at (11.4,-1.8) {$\tilde{p}_2,p_2$};
\node[circle, draw, minimum height=8mm](C) at (12,0){$2$};
\node[] at (14.5,0) {$\dots$};
\node[circle, draw, minimum height=8mm](D) at (17,0){$2$};
\node[circle, draw, minimum height=8mm](C1) at (20.7,-2.5){$1$};
\node[] (H3) at (18,1.8) {$\tilde{q}_1,q_1$};
\node[] (H4) at (18,-1.8) {$\tilde{q}_2,q_2$};
\node[rectangle, draw,  minimum width=8mm, minimum height=8mm](D1) at (20.7,2.5){ $1$};

\draw[-] (C) -- (B);
\draw[-] (A) -- (C);
\draw[-] (C1) -- (D);
\draw[-] (D) -- (D1);
\end{tikzpicture}
\end{center} 
where we have relabelled $b_2$, $\tilde{b}_2$ as $q_i$, $\tilde{q}_i$. In total we have k gauge groups in the quiver and the superpotential reduces precisely to that of the $\CN=4$ theory, therefore making the enhancement of supersymmetry manifest. This is indeed the known mirror dual of $SU(2)$ SQCD \cite{Intriligator:1996ex}. The case of $U(2)$ SQCD can be treated in the same way. 

The case $k=3$ deserves some further comments since in this case there is only one $U(2)$ gauge group in the mirror quiver. After the deformation $U(2)$ is broken spontaneously to $U(1)$ as before and we are left with the following dual theory 
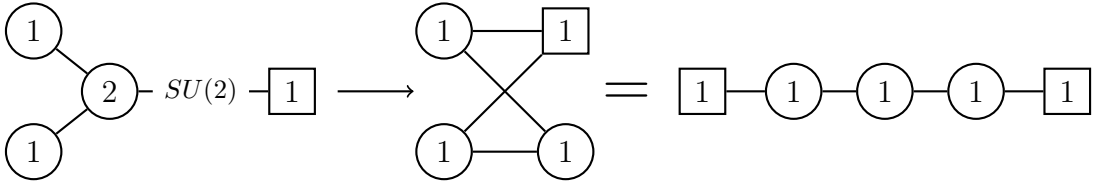
\begin{center}
\begin{tikzpicture}[->,thick, scale=0.4]
\node[circle, draw, minimum height=6mm](A) at (12.5,2){$1$};
\node[circle, draw, minimum height=6mm](B) at (12.5,-2){$1$};
\node[circle, draw, minimum height=6mm](C) at (15,0){$2$};
\node[](C1) at (18,0){{\footnotesize $SU(2)$}};
\node[rectangle, draw,  minimum width=6mm, minimum height=6mm](D1) at (21,0){ $1$};

\node[circle, draw, minimum height=6mm](A2) at (26,2){$1$};
\node[circle, draw, minimum height=6mm](B2) at (26,-2){$1$};
\node[circle, draw, minimum height=6mm](C2) at (30,-2){$1$};
\node[rectangle, draw,  minimum width=6mm, minimum height=6mm](D2) at (30,2){ $1$}; 
\node[](E2) at (32,-0.1){{\huge $=$}}; 

\node[rectangle, draw,  minimum width=6mm, minimum height=6mm](D3) at (34.5,0){ $1$}; 
\node[circle, draw, minimum height=6mm](A3) at (37.5,0){$1$};
\node[circle, draw, minimum height=6mm](B3) at (40.5,0){$1$};
\node[circle, draw, minimum height=6mm](C3) at (43.5,0){$1$};
\node[rectangle, draw,  minimum width=6mm, minimum height=6mm](E3) at (46.5,0){ $1$}; 

\draw[-] (C) -- (B);
\draw[-] (A) -- (C);
\draw[-] (C1) -- (C);
\draw[-] (C1) -- (D1);
\draw[=>] (22.5,0)--(25,0);
\draw[-] (A2) -- (C2);
\draw[-] (A2) -- (D2);
\draw[-] (B2) -- (C2);
\draw[-] (B2) -- (D2);
\draw[-] (D3) -- (A3);
\draw[-] (A3) -- (B3);
\draw[-] (B3) -- (C3);
\draw[-] (C3) -- (E3);
\end{tikzpicture}
\end{center} 
and on the right we recognize the mirror dual of $\CN=4$ SQED with 4 flavors. We thus reach the conclusion that $SU(2)$ SQCD with 3 flavors and SQED with four flavors are equivalent in the infrared. Indeed, the global symmetry is $SU(4)\times U(1)$ in both cases: in the non-abelian theory the $U(1)$ factor arises because the monopole operator has R-charge 1 whereas in the abelian theory it corresponds to the topological symmetry. The monopoles of SQED (whose R-charge is two) are mapped to the dressed monopole and the quadratic Casimir of the adjoint in the $SU(2)$ theory. The meson matrices are mapped to one another. This is in perfect agreement with the brane analysis of \cite{Ferlito:2016grh}. 

Finally, we would like to discuss the case $k=2$, therefore we start from the mirror dual depicted on the right of Figure \ref{mirrsu22} and turn on a superpotential term linear in $X'$. Apart from the fact that now $\Psi$ in (\ref{pots1}) is a diagonal matrix, the analysis is identical to the previous case, at least until (\ref{vevinter}), since we have $\tilde{b}_1b_1=\tilde{p}_1p_1+\tilde{p}_2p_2$. As a result, we now have the following solution  
\be\label{vevinter2}\tilde{p}_1p_1=\left(\begin{array}{cc} 0 & 1-b \\ 0 & 0\end{array}\right);\quad \tilde{p}_2p_2=\left(\begin{array}{cc} 0 & b \\ 0 & 0\end{array}\right),\ee 
where $b$ is a  generic complex parameter. This indeed is not allowed when $\tilde{p}_i,p_i$ transform as doublets of $U(2)$. For a generic value of $b$, the effective low energy theory is described by a free hyper. However, for $b=0,1$ we find an interacting theory: $\CN=4$ SQED with two flavors. This agrees with the finding of \cite{Dey:2017fqs, Assel:2018exy}: there are two singular points on the CB at which the low energy effective action is $T(SU(2))$. At all other points the effective theory is a twisted hypermultiplet. 

The case of $U(2)$ SQCD with two flavors is analogous: The mirror dual is the quiver 
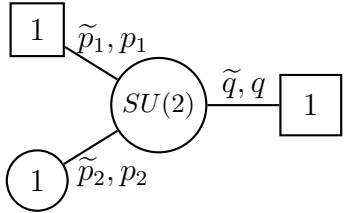
\begin{center}
\begin{tikzpicture}[->,thick, scale=0.4]
\node[rectangle, draw, minimum width=7mm, minimum height=7mm](A) at (9,2.5){$1$};
\node[] (H1) at (11.5,2.2) {$\tilde{p}_1,p_1$};
\node[circle, draw, minimum height=8mm](B) at (9,-2.5){$1$};
\node[] (H1) at (11.5,-2.2) {$\tilde{p}_2,p_2$};
\node[circle, draw, inner sep=2.5](C) at (13,0){{\footnotesize $SU(2)$}};
\node[] (H3) at (15.8,0.7) {$\tilde{q},q$};
\node[rectangle, draw,  minimum width=8mm, minimum height=8mm](D) at (18,0){ $1$};

\draw[-] (C) -- (B);
\draw[-] (A) -- (C);
\draw[-] (C) -- (D);
\end{tikzpicture}
\end{center}
with superpotential (\ref{supcomp}) (now without the singlet $S$, since we are interested in making contact with the $\CN=4$ theory) 
$$\W=X'\tilde{q}q+\epsilon^{\alpha\beta}q_{\alpha}(\phi q)_{\beta}+Y'\epsilon_{\alpha\beta}(\tilde{q}\phi)^{\alpha}\tilde{q}^{\beta}+\tr(\phi\tilde{p}_ip^i)+\varphi p_2\tilde{p}_2.$$
Once we have turned on the linear term in $X'$, we have again the solution (\ref{vevinter2}) and for generic values of $b$ the low energy theory is described by two free hypermultiplets. For $b=0$ we find a singular point with effective theory $T(SU(2))$ plus a free hyper. For $b=1$ instead the effective theory is just given by two free hypermultiplets. Our findings are in perfect agreement with those of \cite{Assel:2017jgo}.

\subsection{Deformation to pure $\CN=2$ SQCD}

In order to flow to $\CN=2$ SQCD we should turn on a mass term for the adjoint chiral. In the mirror theory this is mapped to a superpotential term linear in $Y'$. The resulting superpotential is\footnote{The coefficient 2 in the linear superpotential term is chosen in order to simplify numerical factors in the following equations.} 
\be\label{pots2}\W=Y'\epsilon_{ab}(\tilde{q}\phi)^a\tilde{q}^b+2Y'+X'\tilde{q}q+\epsilon^{ab}q_a(\phi q)_b+\Tr(\Psi b_1\tilde{b}_1)-\Tr(\phi\tilde{b}_1b_1)+\dots\ee
and the F-term for $Y'$ tells us that $\epsilon_{ab}(\tilde{q}\phi)^a\tilde{q}^b$ acquires an expectation value. The D and F terms are satisfied by 
\be\label{vev2}q=0;\quad \tilde{q}=(\begin{array}{cc} \sqrt{2} & 0 \end{array});\quad \phi=\left(\begin{array}{cc} 0 & 1 \\ 0 & 0\end{array}\right)\ee
Notice that here it is crucial that the rightmost gauge group is $SU(2)$ and not $U(2)$, otherwise the above solution would not satisfy the D-term equation. 

The difference with respect to the previous case is that the vev does not ``propagate" to the bifundamental $\tilde{b}_1, b_1$ and therefore all the $U(2)$ gauge groups remain unbroken. When we expand (\ref{pots2}) around the vev we find the terms 
$$\W=X'\langle\tilde{q}\rangle q+\epsilon^{ab}q_a(\langle\phi\rangle q)_b+\dots$$
which give mass to $X'$ and all the components of $q$. Also $Y'$ becomes massive and can be integrated out. As for $\phi$, the Cartan component and $\phi_{12}$ recombine with the $SU(2)$ vector multiplets due to the Higgs mechanism. Taking this into account we conclude that in the effective theory $\phi$ (vev plus fluctuations around it) is of the form 
\be\label{adjeff}\phi=\left(\begin{array}{cc}0 & 1 \\ \varphi & 0\end{array}\right).\ee 
In conclusion, the doublets $\tilde{q}$, $q$ and the singlets $X'$, $Y'$ disappear from the low energy spectrum and the effective theory is described by the quiver 
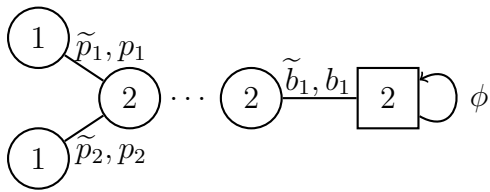
\begin{center}
\begin{tikzpicture}[->,thick, scale=0.4]
\node[circle, draw, minimum height=8mm](A) at (9,2){$1$};
\node[] (H1) at (11.4,1.7) {$\tilde{p}_1,p_1$};
\node[circle, draw, minimum height=8mm](B) at (9,-2){$1$};
\node[] (H1) at (11.4,-1.7) {$\tilde{p}_2,p_2$};
\node[circle, draw, minimum height=8mm](C) at (12,0){$2$};
\node[] at (14,0) {$\dots$};
\node[circle, draw, minimum height=8mm](D) at (16,0){$2$};
\node[] (H4) at (18.2,0.7) {$\tilde{b}_1,b_1$};
\node[rectangle, draw,  minimum width=8mm, minimum height=8mm](C1) at (20.5,0){ $2$};
\node[] (H5) at (23.5,0) {$\phi$};

  \path[every node/.style={font=\sffamily\small,
  		fill=white,inner sep=1pt}]
(C1) edge [loop, out=-30, in=30, looseness=4] (C1);
\draw[-] (C) -- (B);
\draw[-] (A) -- (C);
\draw[-] (C1) -- (D);
\end{tikzpicture}
\end{center}
where $\phi$ is as in (\ref{adjeff}) and is coupled to $\tilde{b}_1, b_1$ (which are now identified with two doublets of $U(2)$) through the superpotential term $\Tr(\phi\tilde{b}_1b_1)$ (see (\ref{pots2})). Notice that this is precisely the theory we end up with if we start from the $\CN=4$ quiver 
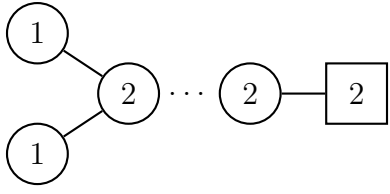
\begin{center}
\begin{tikzpicture}[->,thick, scale=0.4]
\node[circle, draw, minimum height=8mm](A) at (9,2){$1$};
\node[circle, draw, minimum height=8mm](B) at (9,-2){$1$};
\node[circle, draw, minimum height=8mm](C) at (12,0){$2$};
\node[] at (14,0) {$\dots$};
\node[circle, draw, minimum height=8mm](D) at (16,0){$2$};
\node[rectangle, draw,  minimum width=8mm, minimum height=8mm](C1) at (19.5,0){ $2$};

\draw[-] (C) -- (B);
\draw[-] (A) -- (C);
\draw[-] (C1) -- (D);
\end{tikzpicture}
\end{center}  
and then we deform the theory by introducing a chiral multiplet in the adjoint of the $SU(2)$ symmetry (rotating the two flavors on the right) coupled to the corresponding moment map via a superpotential term and turn on a nilpotent vev for it. This is precisely the prescription  proposed in \cite{Giacomelli:2017vgk} to construct the mirror dual of $\CN=2$ $SU(2)$ SQCD. 

The case of SQCD with two flavors deserves some further comments. We start from the quiver discussed in Section \ref{sect1} (see Figure \ref{mirrsu22}) and turn on a superpotential term linear in $Y'$. As a result of this deformation, the $SU(2)$ gauge group is higgsed and we are left with two copies of SQED with two flavors, coupled together via a superpotential term. If we denote by $p_i$ and $q_i$ respectively the hypers charged under the two $U(1)$ gauge groups, we have the superpotential 
\be\W=\alpha\tilde{p}_ip^i+\beta\tilde{q}_iq^i+\tilde{p}_1p^2+\tilde{q}_1q^2+\varphi(\tilde{p}_2p^1+\tilde{q}_2q^1),\ee 
where we have used (\ref{adjeff}). After integrating out massive fields, we are left with two copies of SQED with one flavor coupled together. The superpotential is 
\be\label{effpp}\W=\alpha^2\tilde{p}p+\beta^2\tilde{q}q+\varphi(\tilde{p}p+\tilde{q}q)\ee 
and, by using the duality between SQED with one flavor and $XYZ$ \cite{Aharony:1997bx}, we find a WZ model with 9 chirals (some of them are massive) and superpotential 
\be\label{sup2}\W=xyz+x'y'z'+\varphi(x+x')+\alpha^2x+\beta^2x'\longrightarrow x(yz-y'z'+\alpha^2-\beta^2),\ee 
where we have integrated out massive fields. If we now identify $x$ with the monopole operator of SQCD (as is implied by the chiral ring map discussed in Section \ref{sect1}) and we identify the $4\times4$ antisymmetric meson matrix $M_{ij}$ of SQCD with 
$$\left(\begin{array}{cccc}0 & \alpha+\beta & z' & y \\
-\alpha-\beta & 0 & z & y' \\
-z' & -z & 0 & \alpha-\beta \\
-y & -y' & \beta-\alpha & 0\end{array}\right)$$ 
then (\ref{sup2}) is equivalent to 
$$\W=\mathfrak{M}Pf(M),$$ 
which is known to be the effective low-energy superpotential of $SU(2)$ SQCD with two flavors \cite{Aharony:1997bx}. This provides another test of our duality.

The theory with gauge group $U(2)$ and two flavors can be analyzed in the same way: We have again (\ref{effpp}) but now the theory has just $U(1)$ gauge symmetry instead of $U(1)^2$ (under which, say, only $\tilde{p}$ and $p$ are charged whereas $\tilde{q}$ and $q$ are neutral and play the role of singlets). Using again the duality with $XYZ$ we find a WZ model with superpotential 
\be\W=xyz+\alpha^2x+\beta^2\tilde{q}q+\varphi(x+\tilde{q}q).\ee 
When we integrate out $\varphi$ and $x$ we are left with 
\be\W=\tilde{q}q(\beta^2-yz-\alpha^2),\ee 
which agrees with the effective superpotential $\W=\mathfrak{M}^+\mathfrak{M}^-\det(M)$ describing the moduli space of the theory (see e.g. \cite{Aharony:1997gp}), if we identify $\tilde{q}$ and $q$ with the monopole operators of $U(2)$ SQCD and the meson matrix $M$ with 
$$M\equiv \left(\begin{array}{cc}
\beta+\alpha & y \\
z & \beta-\alpha
\end{array}\right).$$

\bibliographystyle{ytphys}

\end{document}